\documentclass[aps,pre,onecolumn,superscriptaddress,showpacs]{revtex4-1}
\usepackage[english]{babel}
\usepackage[utf8x]{inputenc}
\usepackage{amsmath,amsfonts,amssymb,amsbsy,eucal,dsfont}
\usepackage{natbib}
\usepackage[usenames, dvipsnames]{color}
\usepackage[colorlinks]{hyperref}
\usepackage{graphicx}
\usepackage{rotating}
\usepackage{ulem}
\usepackage{multirow}
\usepackage{longtable}
\newcommand{\hide}[1]{}

\begin{document}

\title{Short- and long-time diffusion, and dynamic scaling in suspensions of charged colloidal particles}

\author{Adolfo J. Banchio}
\affiliation{FaMAF--UNC and IFEG-CONICET, Ciudad Universitaria, 5000 C{\a 'o}rdoba, Argentina}

\author{Marco Heinen}
\affiliation{División de Ciencias e Ingenierías, Universidad de Guanajuato, 37150 León, Guanajuato, Mexico}

\author{Peter Holmqvist}
\affiliation{Division of Physical Chemistry, Lund University, Lund SE-221 00, Sweden}

\author{Gerhard N\"agele}
\email[]{g.naegele@fz-juelich.de}
\affiliation{Institut f\"{u}r Theoretische Physik II, Weiche Materie, Heinrich-Heine-Universit\"{a}t D\"{u}sseldorf, D-40225 D\"{u}sseldorf, Germany}
\affiliation{Institute of Complex Systems (ICS-3), Forschungszentrum J\"ulich, D-52425 J\"ulich, Germany}

\date{\today}

%%%%%%%%%%%%%%%%%%%%%%%%%%%%%%%%%%%%%%%%%%%%%%%%%%%%%%%%%%%%%%%%%%%%%%%%%
\renewcommand{\figurename}{Fig.}
\renewcommand{\figuresname}{Figs.}
\renewcommand{\refname}{Ref.}
\newcommand{\refsname}{Refs.}
\newcommand{\expressionname}{Eq.}
\newcommand{\expressionsname}{Eqs.}
\newcommand{\sectionname}{Sec.}
%%%%%%%%%%%%%%%%%%%%%%%%%%%%%%%%%%%%%%%%%%%%%%%%%%%%%%%%%%%%%%%%%%%%%%%%%

\begin{abstract}
We report on a comprehensive theory-simulation-experimental study of collective and self-diffusion in concentrated
suspensions of charge-stabilized colloidal spheres. In theory and simulation, the spheres are assumed to interact directly by a hard-core plus
screened Coulomb effective pair potential.
The intermediate scattering function, $f_c(q,t)$, is calculated by
elaborate accelerated Stokesian Dynamics (ASD) simulations for Brownian systems where many-particle hydrodynamic interactions (HIs) are fully 
accounted for, using a novel extrapolation scheme to a macroscopically large system size  
valid for all correlation times.
The study spans the correlation time range from the colloidal short-time to the long-time regime.
Additionally, Brownian Dynamics (BD) simulation and mode-coupling theory (MCT) results of $f_c(q,t)$ are generated where
HIs are neglected. Using these results, the influence of HIs on collective and self-diffusion, and the accuracy of the MCT method are quantified. 
It is shown that HIs enhance collective and self-diffusion at intermediate and long times, whereas at short times self-diffusion,
and for certain wavenumbers also collective diffusion are slowed down. 
MCT significantly overestimates the slowing influence of dynamic particle caging.  The dynamic scattering functions obtained in the
ASD simulations are in decent agreement with our dynamic light scattering (DLS) results for a concentration series of
charged silica spheres in an organic solvent mixture, in the experimental time window and wavenumber range.  
From the simulation data for the time derivative of the width function associated with $f_c(q,t)$, 
there is indication of long-time exponential decay of $f_c(q,t)$,  
for wavenumbers around the location of the static structure factor principal peak. 
The experimental scattering functions in the probed time range are consistent with a time-wavenumber factorization scaling behavior of $f_c(q,t)$ that was first reported by Segr\`e and Pusey [Phys. Rev. Lett.{\bf 77}, 771 (1996)] for suspensions of hard spheres.    
Our BD simulation and MCT results predict a significant 
violation of exact factorization scaling which, however, 
is approximately restored according to the ASD results when HIs are accounted for, consistent with the experimental findings for $f_c(q,t)$.  
Our study of collective diffusion is amended by simulation and theoretical results for the self-intermediate scattering
function, $f_s(q,t)$, 
and its non-Gaussian parameter $\alpha_2(t)$, 
and for the particle mean squared displacement 
$W(t)$ and its time derivative. 
Since self-diffusion properties are not assessed in standard DLS measurements, 
a method to deduce $W(t)$ approximately 
from $f_c(q,t)$ is theoretically validated.    
\end{abstract}


\maketitle

%%%%%%%%%%%%%%%%%%%%%%%%%%%%%%%%%%%%%%%%%%%%%%%%%%%%%%%%%%%%%%%%%%%%%%%%%%%%%%%%%%%%%%%%%%%%%%%%%%%%%%%%%%%%%%%%
\section{Introduction}\label{sec:Intro}

Suspensions of Brownian spheres interacting electrostatically by a screened Coulomb effective pair potential \cite{RusselBook:1989,Nagele1996,Likos_MicrogelReview:2011,Holmqvist2012,ivlev2012complex,NaegeleEPJST:2013}  
are a paradigm
for a vast variety of charge-stabilized globular colloids encountered in chemical industry, biology and medicine. 
The dynamics of the interacting particles can be conveniently probed for an extended range of scattering wavenumbers $q$ and correlation
times $t$, using dynamic light scattering (DLS) \cite{SegrePusey1996,Loewen1993,Holmqvist2010}, and more recently also using x-ray photon correlation spectroscopy (XPCS)  (see, e.g., \cite{Lurio2000,Westermeier2012}).
The compatibility of the two methods and their respective advantages and disadvantages have been thoroughly assessed in a recent study by Martinez {\it et al}. \cite{Martinez2011}.

In addition to electro-steric forces, the dynamics of the spheres is significantly affected by the solvent-mediated hydrodynamic interactions (HIs) that are in general long ranged, and non-pairwise additive at larger particle concentrations \cite{NaegeleVarena2013}. This poses a challenge to computer simulation studies,
and to the development of analytic theoretical schemes, in particular when dynamic scattering functions at intermediate and long times are explored.

The fundamental quantity determined in DLS and XPCS experiments as a function of scattering wavenumber $q$ and correlation time $t$, is the normalized intermediate scattering function
\begin{equation}
\label{eq:fct}
  f_c(q, t) = \frac{S(q,t)}{S(q)}= \exp\{-q^2\;\!w_c(q,t)\} \,.
\end{equation}
Here, $w_c(q,t)$ is the associated collective width function \cite{Martinez2011}, 
and $S(q,t)$ denotes the dynamic structure factor whose initial value is equal to
the static structure factor $S(q)$. In the colloidal short-time regime characterized by $\tau_B\ll t\ll \tau_a$,
the function $f_c(q,t)$ decays exponentially in time according to \cite{Pusey1990}
\begin{equation}
\label{eq:fct-shorttime}
  f_c(q, t) \approx \exp\{-q^2\!\;D_s(q)\;\!t\} \,,
\end{equation}
where
\begin{equation}
\label{eq:Dsq}
  D_s(q) = d_0 \;\!\frac{H(q)}{S(q)}
\end{equation}
is the wavenumber-dependent short-time diffusion function proportional to the hydrodynamic function $H(q)$.
The symbol $d_0$ denotes the Stokes-Einstein single-sphere translational free diffusion coefficient, $\tau_a=a^2/d_0$ is the associated characteristic time for diffusion 
across a distance equal to the particle hydrodynamic radius $a$, and $\tau_B$ is the characteristic decay time of particle momentum auto-correlations. Owing to solvent friction, $\tau_B$ is several orders of magnitude smaller than $\tau_a$ \cite{Nagele1996}.
The hydrodynamic function plays the role of a generalized short-time sedimentation coefficient \cite{LaddWeitz:1995,Abade2010}, 
and it directly reflects the influence of the HIs. At infinite dilution, or in the (hypothetical) absence of HIs, $H(q)$ is identically equal to one. 
The HIs give rise to undulations in $H(q)$ at non-zero particle concentration. For large wavenumbers $q$ well above the value, 
$q_m$, where the principal peak of $S(q)$ is located, $H(q)$ becomes equal to the normalized short-time self-diffusion coefficient $d_s/d_0$, with $d_s$ being the initial slope of the particle mean squared displacement (MSD) in three dimensions,
\begin{equation}
\label{eq:MSD}
  W(t) = {\Big<} [{\bf r}(t)-{\bf r}(0)]^2 {\Big>}/6 \,,
\end{equation}
were ${\bf r}(t)$ is a particle center position at time $t$ and the brackets denote an equilibrium ensemble average.
The influence of the near-distance part of the HIs manifests itself in values of $d_s$ smaller than $d_0$.
The MSD $W(t)$ increases sublinearly for intermediate times $t \sim \tau_a$. At long times, it increases linearly with its slope equal to the long-time self-diffusion coefficient $d_l$ that in general is smaller than $d_s$.
Note that $D_s(q)$ is equal to the initial (i.e. short-time) slope of $w_c(q,t)$.

At longer times $t \gtrsim \tau_a$, the strictly monotonic time decay of $f_c(q,t)$ becomes slower than that 
described by the short-time exponential form in Eq. (\ref{eq:Dsq}), owing to the dynamic caging of the spheres by neighboring ones. 
The decay of $f_c(q,t)$ at intermediate and especially long times (where $t \gg \tau_a$ in the latter case)
is difficult to describe theoretically for non-zero $q$ values and larger concentrations, since the concerted influence of direct interactions and HIs on microstructural changes across distances $\sim 2\pi/q$ needs to be accounted for.
This is why for charge-stabilized suspensions only few simulation results for $f_c(q,t)$ are reported extending well beyond the short-time regime. In most of these earlier simulations, HIs are computed with strong approximations \cite{vanMegenSnook1988,NaegeleMolphys:2002}, or disregarded
altogether \cite{Haertl1992,Loewen1993,WagnerHaertl:2001}. It would be thus very useful if the long-time dynamics could
be related to the exponential short-time decay of $f_c(q,t)$, since regarding the latter accurate semi-analytic methods
of calculating $H(q)$ and $S(q)$ in Eq. (\ref{eq:fct-shorttime}) have been developed \cite{Banchio2008,Heinen2010,Heinen2011_dyn,Heinen2011}.

Regarding suspensions of electrically neutral colloidal hard spheres,
such a relation was proposed empirically by Segr\`e and Pusey \cite{SegrePusey1996,SegrePusey1997,PuseyPoon1997}
on basis of their two-color DLS measurements of $S(q,t)$ covering the fluid-phase concentration region.
They found that the collective width function is well approximated for wavenumbers $q/q_m \geq 0.7$ including the principal peak
region of $S(q)$, during their probed correlation time window, by the expression
\begin{equation}
\label{eq:PuesySegreScaling}
  w_c(q,t) \approx  \frac{D_s(q)}{d_s}\;\!W(t) \,.
\end{equation}
In this so-called Segr\`e-Pusey (or time-wavenumber) factorization scaling expression, the wavenumber and time dependencies of the width function are factorized.
Very remarkably, the time dependence is predicted here to be solely determined by self-diffusion in terms of the MSD,
and the wavenumber dependence of $w_c(q,t)$ is embodied solely in the short-time diffusion function $D_s(q)$.
At short times where $W(t) \approx d_s t$ applies, the above scaling expression for $w_c(q,t)$ reproduces the exact short-time form of $f_c(q,t)$ given in
Eq. (\ref{eq:fct-shorttime}). Note that the factorization scaling expression in Eq. (\ref{eq:PuesySegreScaling}) is based on two interrelated assumptions: 
The first one claims the existence of an exponentially decaying long-time mode of $f_c(q,t)$, since $W(t) = d_l\;\!t$ for $t \gg \tau_a$.
The second and more constraining assumption is the wavenumber-time factorization of the collective width function at all times, implying for long times in particular that
\begin{equation}
 \label{eq:PuseySegreScaling-long}
  w_c(q,t\gg \tau_a) =  D_l(q)\;\!t \,,
\end{equation}
with a long-time diffusion function
\begin{equation}
 \label{eq:Dlq}
  D_l(q) \approx  D_s(q)\;\!\frac{d_l}{d_s} \,.
\end{equation}
This long-time function has the same wavenumber dependence as its short-time counterpart $D_s(q)$, up to a constant scale factor smaller than one given by the ratio of
long-time and short-time self-diffusion coefficients. Stated alternatively, valid Segr\`e-Pusey factorization scaling implies the ratio
\begin{equation}
 \label{eq:width-ratio}
  \frac{w_c(q,t)}{D_s(q)\!\;t} \approx \frac{W(t)}{d_s\;\!t} \,,
\end{equation}
of the width function and its short-time form to be wavenumber-independent,
with its time dependence determined for all times (i.e., not only for $t \ll \tau_a$) 
by the MSD divided by $d_s t$.
Consequently, the ratio on the left-hand-side of Eq. (\ref{eq:width-ratio}) should converge to $d_l/d_s$ in the long-time limit.

An exponential long-time mode, as implied by the factorization scaling expression for $t \gg \tau_a$ 
and $q/q_m \geq 0.7$, is an intricate feature which is hard to justify theoretically,
and difficult to discern empirically from the noise background in scattering experiments and simulations (see later).
Moreover, the possible existence of a long-time exponential mode of $f_c(q,t)$ does not necessarily imply the validity of the wavenumber-time factorization in $w_c(q,t)$ according to Eq. (\ref{eq:PuesySegreScaling}). 
In fact, a long-time exponential mode is obtained in the hydrodynamic (Markovian) limit of $q\to 0$ and $t\to \infty$ with $q^2\,\!t$ kept constant,
where the collective width function becomes equal to $d_l^c\;\!t$.
Here, $d_l^c$ is the gradient or long-time collective diffusion coefficient associated with density fluctuations of macroscopically large wavelengths, 
and  the transport coefficient appearing in Fick's constitutive law for the macroscopic diffusion current.
However, the small-$q$ regime is outside the supposed application range of factorization scaling.

In the opposite limit of large wavenumbers where small distances are probed and for which $S(q) \approx 1$ and $H(q) \approx d_s/d_0$ are valid,
a valid factorization scaling implies
\begin{equation}
 \label{eq:fc-large-q}
  f_c(q\gg q_m,t) \approx \exp\{-q^2\;\!W(t)\} \,.
\end{equation}
At large-$q$ values, $f_c(q,t)$ reduces to the self-intermediate scattering function \cite{Nagele1996},
\begin{equation}
 \label{eq:fsq}
  f_s(q,t) = \exp\{-q^2\;\!w_s(q,t)\} \,,
\end{equation}
related to colloidal self-diffusion.
The associated self-diffusion width function, $w_s(q,t)$, would be $q$-independent
and equal to $W(t)$ only if non-Gaussian contributions were negligibly small \cite{vanMegenUnderwood:1989,Nagele1996,vanMegenSchoepe:2017}.
For larger concentrations, this holds for small wavenumbers only, of values well below those for which 
factorization scaling is supposed to apply. In this context, note that $w_s(q\to 0,t)=W(t)$, and that non-Gaussian contributions 
causing $w_s(q,t)$ to differ from $W(t)$ and to be $q$-dependent are small for times $t$ where $q^2\;\!W(t) \lesssim 1$ \cite{vanMegenUnderwood:1989,Brands:1999}.

We emphasize that the genuine long-time regime of $f_c(q,t)$ is in general not resolved in scattering experiments 
since it is commonly located in the noise floor of the scattering function for which $t \gg 1/(q^2\;\!d_0)$. Likewise, the genuine long-time regime is difficult to access
in simulations when $f_c(q,t)$ is directly considered. Therefore, $D_l(q)$ can be defined and assessed in this direct way only in an operational sense in experiment and simulation, 
by inspecting whether for the largest accessed correlation times a plateau region of the width
ratio $w_c(q,t)/(D_s(q)\;\!t)$ is reached or approached. The plateau value is then identified 
operationally as the ratio $D_l(q)/D_s(q)$. If additionally to the long-time exponential decay also the other features of factorization
scaling are valid, $D_l(q)/D_s(q)$ should be $q$-independent and equal to $d_l/d_s$. We will show that it is proficient to deduce $D_l(q)$ from the plateau region of $d w_c(q,t)/dt$ since the plateau region for the time derivative of $w_c(q,t)$ is reached more quickly than that for $w_c(q,t)/(d_s t)$.

While Segr\`e and Pusey found $D_l(q)/D_s(q)$ to be practically $q$-independent in a broad wavenumber range
including the principal peak region of the hard-sphere $S(q)$, their finding conflicts with  
a later XPCS study of Lurio {\it et al}. \cite{Lurio2000} on charged polystyrene spheres in glycerol (see also \cite{Lumma2000}).
The $S(q)$'s determined by Lurio {\it et al.} are indistinguishable from those of hard spheres,
but their measured ratio $D_l(q)/D_s(q)$ varies markedly as a function of $q$ for larger values of $\phi$.

From DLS measurements on charge-stabilized silica spheres in a low-salinity index-matching $80:20$ toluene:ethanol solvent mixture, Holmqvist and N\"agele \cite{Holmqvist2010} found factorization scaling to be approximately valid in the experimental time window, but only within a wavenumber interval enclosing the structure factor peak position, $q_m$, distinctly narrower than that identified by Segr\`e and Pusey for hard spheres.
The DLS measurements in \cite{Holmqvist2010} cover the volume fraction range up the freezing transition value.
Quite remarkably, the self-diffusion coefficients ratio, $d_l/d_s$, at freezing concentration 
deduced from the scattering data on assuming validity of factorization scaling is close to the characteristic value $0.1$ 
of the L\"owen-Palberg-Simon dynamic freezing criterion \cite{Loewen1993}.
The primarily experimental study by Holmqvist and N\"agele includes also a brief theoretical analysis pointing to the
approximate nature of factorization scaling, and to the problem of establishing theoretically the existence of a true exponential long-time decay of $f_c(q,t)$. 

More recently, Martinez {\it et al.} \cite{Martinez2011} combined DLS and XPCS to investigate the validity of factorization scaling in a suspension of sterically stabilized methacrylate spheres in cis-decalin.
They showed that both scattering methods give the same results for $f_c(q,t)$ in the overlapping range of wavenumbers and correlation times, eliminating thus one possible reason for the conflicting hard-sphere suspension results of Lurio {\it et al}. and Segr\`e and Pusey. 
For large concentrations $\phi\sim 0.4-0.5$, their scattering data are in accord with the factorization scaling of $w_c(q,t)/(D_s(q)\;\!t)$ for
several orders of magnitude in the correlation time. However, Martinez {\it et al.} do not observe a genuine long-time exponential decay at wavenumbers different from $q_m$.
In contrast to what is reported in \cite{Martinez2011}, more recent DLS experiments by Joshi {\it et al.} \cite{Joshi:2013} on aqueous dispersions
of thermo-responsive PNIPAM ionic nanogel particles are again compatible with a $q$-independent ratio $D_l(q)/D_s(q)$. DLS experiments on an additional suspension of soft particles, namely soft giant micelles, are likewise compatible with a long-time exponential decay at the structural peak position wavenumber $q_m$ \cite{Sigel:1999}. 

Theoretical and simulation works on long-time collective diffusion in general, and on the validity of factorization scaling in particular, are scarce and mainly deal with hard-sphere suspensions without the salient
HIs included. Most notable is here the theoretical work by Cichocki and Felderhof where $f_c(q,t)$ and its spectral density function are calculated for a semi-dilute hard-sphere suspension \cite{CichockiFelderhof:1993}. These authors have further derived a so-called contact Enskog type approximation (CEA) scheme for the hard-sphere $f_c(q,t)$ without HIs that is applicable up to moderately large concentrations \cite{CichockiFelderhof:1994}. According to their analysis in which HIs are disregarded, $f_c(q,t)$ shows indeed a genuine long-time exponential decay for non-small $\phi$, and for wavenumbers where $S(q)$ is sufficiently large according to some theoretical criterion \cite{CichockiFelderhof:1994}, but $D_l(q)/D_s(q)$ is predicted to vary significantly as a function of $q$. We mention here also related MCT-based work on hard-sphere suspensions
by Fuchs and Mayr \cite{FuchsMayr:1999} where the $\alpha$-relaxation behavior of $f_c(q,t)$ in the supercooled regime near the glass transition volume fraction was studied. They find factorization scaling to hold approximately in the $\alpha$ scaling time window except for small wavenumbers. Regarding theoretical studies of the dynamics of charge-stabilized suspensions, a few MCT results for the concentration dependence of $D_l(q_m)$ were discussed in \cite{Banchio:2000,NaegeleMolphys:2002}, but wavenumbers different from $q_m$ and HIs were not considered.

The above survey on conflicting findings related to the validity of factorization scaling amply demonstrates the high demand on 
simulation-theoretical work in order to improve our understanding of many-particle diffusion of interacting colloidal particles at
intermediate and long times, and to quantify the influence of the HIs. The present paper copes with this demand. Since factorization scaling relates collective to self-dynamics, and long-time to short-time properties, we study both the behavior of $f_c(q,t)$ and $f_s(q,t)$, and of the associated width functions. Self-diffusion properties were not determined in the
aforementioned experimental studies dealing with the factorization scaling conundrum. We report here on our 
elaborate ASD and BD simulation study of the intermediate and self-intermediate functions and their width functions, 
and of $W(t)$, $D_s(q)$ and the non-Gaussian parameter $\alpha_2(t)$, for fluid-phase suspensions of charge-stabilized spheres at three different concentrations.
We are concerned here with charge-stabilized suspensions of low salinity, in which the colloidal particles and their microstructure are distinctly different from those of hard spheres due to the long-ranged electrostatic repulsion. 
A comprehensive analysis of the dynamics in concentrated suspensions of neutral hard spheres constituting the high-salinity limit of charged particles suspensions (except for residual van der Waals attraction)
will be presented in a forthcoming publication.

Due to the long-range nature of the HIs, a finite system size correction procedure is required to extrapolate the ASD data to an infinitely large (macroscopic) suspension.
For this purpose, we introduce a generalized finite-size correction scheme for $f_c(q,t)$ 
and the MSD $W(t)$ that applies to arbitrary correlation times.
Moreover, our simulation data are used to scrutinize an approximate procedure, proposed by Pusey \cite{Pusey1978, Abade2010}, to extract the self-diffusion
coefficients $d_s$ and $d_l$, and to a certain accuracy also $f_s(q^\ast,t)$ and $W(t)$, from DLS measurements performed at a 
wavenumber $q^\ast >q_m$ where $S(q^\ast)=1$.
The self-intermediate scattering function $f_s(q,t)$ at arbitrary $q$, and hence $W(t)$, have been determined experimentally
only by application of the DLS technique to a special binary mixture consisting of a host and a trace component of particles of practically equal size and effective charge,
where the host particles are refractive index-matched to the solvent. Such mixtures have been studied in the seminal works of van Megen and collaborators,
both for neutral hard spheres \cite{vanMegenUnderwood:1989} and charge-stabilized particles \cite{Brands:1999}. 

Our simulation results are amended by MCT calculations of collective and self-diffusion properties that span a much broader time range than the one accessible in the simulations. The presented ASD simulations account for the many-body HIs which makes them particularly
costly in terms of computation time, in particular since the long-range HIs necessitate a size extrapolation 
to an infinitely large system.  The significant influence of the HIs is assessed from comparing the ASD results with our corresponding BD simulation results where HIs are disregarded. The BD data, in turn, are used to scrutinize the accuracy of MCT calculations that  
are computationally fast in comparison. The MCT scheme requires $S(q)$ as its only input which is computed  using the numerically efficient so-called modified penetrating-background corrected rescaled mean spherical approximation (MPB-RMSA) integral equation method \cite{Heinen2011}. In addition, we have performed Monte-Carlo (MC) simulations of $S(q)$,  first to show the high accuracy
of the MPB-RMSA results for the explored systems, and second to profitably use the MC generated equilibrium particle configurations in our
ASD simulation calculations of the short-time diffusion properties $H(q)$ and $d_s$ appearing in the factorization scaling expression.
The hydrodynamic function, $H(q)$, is also calculated using the computationally efficient $\delta\gamma$ scheme of Beenakker and Mazur
with a self-part correction included \cite{Beenakker1983,Mazur1984,Heinen2011_dyn}. This semi-analytic method uses as its input the pair
distribution functions obtained from Fourier-inverting the MPB-RMSA solution for $S(q)$. We show that the theoretical and simulated $S(q)$ and $H(q)$ are in very good agreement with the experimental ones, in the whole experimentally accessed wavenumber range.

The effective colloidal particle pair potential, $u(r)$, used in our simulations and analytical-theoretical calculations is the standard repulsive hard-sphere plus
screened Coulomb potential of Derjaguin-Landau-Verwey-Overbeek (DLVO) type \cite{Verwey_Overbeek1948},
\begin{equation}
 \label{eq:DLVO}
  \beta\;\! u(r) =
  \left\lbrace
  \begin{array}{ll}
  L_B\;\!Z^2\;\!\left[\dfrac{\exp(\kappa a)}{1+\kappa a}\right]^2 \;\!\dfrac{\exp\{-\kappa\;\!r\}}{r}\,,&\quad r>\sigma \,,\\
  ~\\
  \infty, &\quad r \leq \sigma,
  \end{array}
  \right.
\end{equation}
referred to for short as the hard-sphere plus Yukawa (HSY) potential.
In Eq.~\eqref{eq:DLVO}, the electrostatic screening parameter, $\kappa$,
is given in units of the sphere diameter $\sigma = 2a$ by
\begin{equation}
 \label{eq:kappa}
  \left(\kappa\;\! \sigma\right)^2 = \frac{8\;\!L_B/\sigma}{1-\phi}\left[3\;\!\phi |Z| + \pi\;\!n_s\;\!\sigma^3 \right] \,.
\end{equation}
Here, $n_s$ is the number concentration of added $1-1$ electrolyte, $\phi=\left(4\pi/3\right)n a^3$ is the volume fraction
of spheres of number concentration $n$, and $L_B=e^2/(\epsilon k_B T)$ is the Bjerrum length of the
suspending structureless Newtonian fluid of dielectric constant $\epsilon$ at absolute temperature $T$. Moreover, $e$ is the proton elementary charge
and $\beta=1/(k_B T)$ is the inverse thermal energy in terms of the Boltzmann constant $k_B$. In Eq. (\ref{eq:kappa}), it is assumed that the counterions released from the 
sphere surfaces are monovalent. The factor $(1-\phi)$ in the denominator corrects for the volume accessible to the
microions that can not penetrate the colloidal spheres \cite{Russel1981,Denton2000}. The parameter $Z$ is the effective particle charge in units
of $e$ which is smaller in general than the bare surface charge since it accounts for the quasi-condensation of counterions
at the colloid surfaces. The dependence of $Z$ on the bare charge, and on $\kappa \sigma$, $L_B/\sigma$ and $\phi$ in  concentrated
suspensions is the subject of ongoing research
\cite{Alexander1984, HansenLoewen2000, Trizac2003, Pianegonda2007, Torres2008, Colla2009, dosSantos2010, Denton2010, Heinen2014, Boon2015}, 
but not relevant for the present study.

The HSY potential captures essential features of charge-stabilized sphere suspensions, for conditions where the
short-range van der Waals inter-particle attraction can be neglected \cite{Gruijthuisen:2013}. 
These conditions are met for the here considered concentration series of
fluid suspensions of negatively charged trimethoxysilylpropyl methacrylate (TPM) coated silica spheres of mean
radius $\overline{a}=136$ nm in a low-salinity index-matched  $80:20$ toluene:ethanol mixture ($L_B=8.64$ nm at $T=293$ K)
which we study in this work using DLS and static light scattering (SLS). Part of the experimental data for
this system has been published already in \cite{Holmqvist2010,Heinen2010} and are reused here
(as indicated in the respective figures) for the comparison with our new simulation and MCT results, and to allow for a 
complete discussion of the explored concentration series of systems. For the here studied silica spheres systems,
the fluid to fcc crystal freezing transition takes place at $\phi \approx 0.16$ \cite{Holmqvist2010}.

The paper is organized as follows:
In Sec. \ref{sec:Experiments}, the details of our experiments are discussed, including particle synthesis and 
particle characterization
by SAXS, SLS and DLS measurements. A description of the analytical-theoretical methods used for calculating
static and diffusion properties is given in Sec. \ref{sec:Theory}. Sec. \ref{sec:Simulations} provides the essentials
of the employed MC, BD and ASD simulation methods. In Subsec. \ref{sec:sub:ASD} describing the ASD method,
we present in particular a generalization of the finite system size correction method for $f_c(q,t)$ and $f_s(q,t)$ to intermediate and long times. 
Our simulation, theory and experimental results are shown in Sec. \ref{sec:Results}. The first two subsections here are concerned with
statics and short-time diffusion, setting the stage for our discussion, in the subsequent three subsections,
of collective and self-diffusion properties at intermediate to long times. 
A critical assessment is given of the validity of factorization scaling in charge-stabilized suspensions.
Our conclusions in Sec. \ref{sec:Conclusions} are followed by the list of abbreviations.

%%%%%%%%%%%%%%%%%%%%%%%%%%%%%%%%%%%%%%%%%%%%%%%%%%%%%%%%%%%%%%%%%%%%%%%%%%%%%%%%%%%%%%%%%%%%%%%%%%%%%%%%%%%%%%%%
\section{Experimental}\label{sec:Experiments}

\subsection{Charged silica spheres suspension}\label{sec:sub:Synthesis}
The colloidal dispersions studied in this paper consist of TrimethoxysilylPropyl Methacrylate (TPM)-coated
silica spheres in an in index-matched \mbox{$80:20$ toluene:ethanol} solvent mixture. Such particulate suspensions
have first been synthesized and used in light-scattering studies by Philipse and Vrij.
For details on the synthesis procedure, we refer to their original articles in \cite{Philipse1989} and \cite{Philipse1988}.
In suspension, the coated silica particles acquire a negative surface charge via the dissociation of silanol groups \cite{Heinen2014}.
Using small-angle x-ray scattering (SAXS) and a polydisperse form factor model fit to the large-$q$ SAXS intensities,
we have determined the mean particle radius $\overline{a} = 136$ nm, and a rather small size polydispersity
factor (i.e., relative particle diameter standard deviation) of $s = 0.06$. The details of the SAXS form factor fit are discussed in the following. In the subsequent subsection
on static and dynamic light scattering from nearly index-matched silica spheres, where the core-shell structure of the particles is resolved,
a more refined form factor model is used.

\subsection{Small-angle x-ray scattering}\label{sec:sub:SAXS}

The SAXS experiment was performed at the Tro\"{\i}ka III part of the ID10A beamline of the
European Synchrotron Radiation Facility (ESRF) in Grenoble, France.
The experimental details concerning the beam characteristics and the general setup have been described elsewhere
\cite{Thurn-Albrecht2003, Gapinski2009},
and can also be found on the ESRF web page \cite{ESRFwebpage}.
The average current of the beam was around $200$ mA.
We utilized radiation of wavelength $1.55$ \AA{}, corresponding to a photon energy of $7.99$ keV.
The size of the pinhole placed in front of the sample at distance of $25$ cm is $24 \times 24$ $\mu m^2$.
The capillary containing the sample was mounted in a chamber connected to a pipe guiding the scattered x-ray photons to the detector
placed at the distance of 3.480 m from the sample.

In our SAXS measurements, we have measured the mean scattered intensity, $I(q)$, as
a function of the scattering wavenumber $q$. For negligible multiple scattering contributions in a size-polydisperse system, $I(q)$ is given by \cite{Nagele1996}
\begin{equation}
\label{eq:Iq_static}
  I(q) \propto n\;\! \overline{f^2}(q=0)\;\!P_m(q)\;\! S_m(q)\,,
\end{equation}
where $n$ is the colloid mean number density, $\overline{f^2}(q)$ is the second moment of the distribution of field scattering amplitudes $f(q)$, and
$P_m(q)=\overline{f^2}(q)/\overline{f^2}(0)$ is the average (measurable) particle form factor, normalized such that $P_m(q=0)=1$. Here, $S_m(q)$ is the
so-called measurable static structure factor which depends on the size-dependent scattering amplitudes distribution in addition to the pair interactions. 
The optical contrast for x-rays between solvent and core and shell of the spheres is large enough to treat the spheres as
homogeneous. The second moment of the scattering amplitudes distribution is then given by
\begin{equation}\label{eq:Pq}
 \overline{f^2}(q) =  \displaystyle{\int_0^\infty \mathrm{d}a  \, p(a) \, b^2(qa) \,  v(a)^2}.
\end{equation}
with the form amplitude $b(qa) = 3[\sin(qa) - qa \cos(qa)]/(qa)^3$ of a homogeneously scattering sphere of radius $a$
and volume $v(a) = (4\pi/3) a^3$. The size distribution function, $p(a)$, is described here using the skew-symmetric,
unimodal continuous Schulz-Zimm distribution \cite{Schulz1939, Zimm1948},
\begin{equation}\label{eq:Schulz-Zimm}
p(a) = \frac{a^{t}}{\Gamma (t+1)} \, \left( \frac{t+1}{\overline{a}} \right)^{t+1} \,  \exp \left( - \, \frac{t+1}{\overline{a}} \, a \right),
\end{equation}
dependent on the mean particle radius $\overline{a}$, and the polydispersity factor (relative standard deviation) $s = {(t+1)}^{-1/2}$.

\begin{figure}
\centering
\includegraphics[width=0.45\columnwidth]{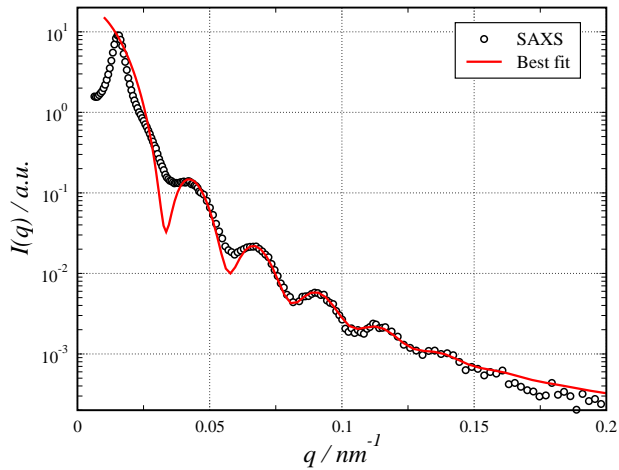}
\vspace{0.0 em}
\caption{\label{fig:Pq}
Black circles: Mean SAXS intensity, $I(q)$, in arbitrary units for a silica-spheres suspension of  volume fraction
$\phi = 0.105$, salt (1-1 electrolyte) concentration $n_s = 0.7$ $\mu$M, and toluene:ethanol solvent at temperature $T=20^o$C.
Red solid curve: Best-fit measurable form factor, $P_m(q)$, according to Eqs. (\ref{eq:Iq_static}) - (\ref{eq:Schulz-Zimm}),
describing homogeneously scattering polydisperse spheres. The form factor fit to the SAXS data in the wavenumber range
$q \gtrsim 0.07$ nm$^{-1}$ gives the mean radius $\overline{a} = 136$ nm, and the size-polydispersity factor
$s=0.06$. The low-$q$ downturn and peak in the SAXS $I(q)$ originate from the corresponding features in $S_m(q)$.
}
\end{figure}

We have determined $\overline{a}$ and $s$ by
fitting the SAXS intensity $I(q)$ for large values of $q$, where $S_m(q) \approx 1$, to the (non-normalized) mean form factor
in Eqs. (\ref{eq:Pq} - \ref{eq:Schulz-Zimm}) describing polydisperse homogeneous spheres. See Fig. \ref{fig:Pq}) for details on the SAXS intensity form factor fitting.

\subsection{Static and dynamic light scattering}\label{sec:sub:SLS_DLS}

The incoherent scattering contribution to $I(q)$ for visible laser light or x-rays is most significant at very small $q$ values located well to the
left of the principal peak position, $q_m$, of $S_m(q)$ \cite{Pusey1982, Nagele1996}. This contribution can considerably complicate the quantitative analysis
of static and dynamic scattering data for systems with pronounced polydispersity such as globular protein solutions
which tend to form clusters of varying sizes \cite{Egelhaaf2004, Heinen2012}.
Yet, in a recent x-ray scattering study by part of the present authors \cite{Westermeier2012},
aqueous suspensions of charged poly-acrylate spheres of mean diameter, effective charge,
and size polydispersity similar to those of the present silica suspensions have been analyzed.
For a small size-polydispersity of about $6\%$ as in the present silica system,
it was shown in this study that the incoherent scattering contribution is negligible
unless one focuses on the low-$q$ behavior of $I(q)$ where, however, the experimental
error bars become largest. Since the experimental low-$q$ regime is not of interest in the present study,
and since polydispersity is small, we can neglect the small size polydispersity in the evaluation of our
SLS and DLS data, and in the theoretical modeling of the pair potential. Different from x-ray scattering,
the core-shell structure of the TPM-coated silica particles does play a role in visible light scattering,
since the particles are nearly refractive index matched to the solvent to minimize multiple light scattering.

\subsection{Polydispersity and form factor}\label{sec:sub:Polydisp}
\begin{figure}
\centering
\includegraphics[width=0.45\columnwidth]{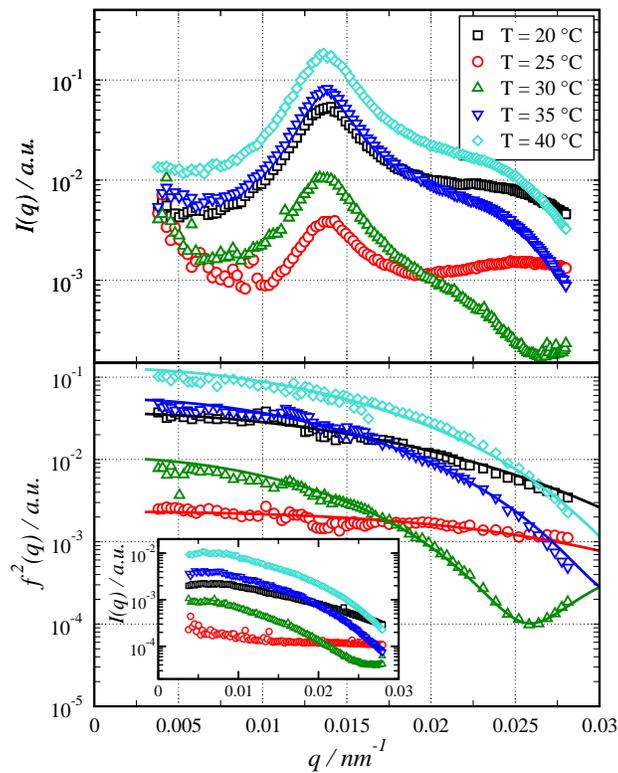}
\vspace{0 em}
\caption{\label{fig:SLS_Pq}
Top panel: Mean SLS intensity, $I(q)$, in arbitrary units, for different
suspension temperatures indicated in the legend in centigrades Celsius.
All results are for a salt concentration of $n_s = 0.7 \mu$M.
Bottom panel: non-normalized best-fit core-shell form factor, $f^2(q)$, according
to Eq. \eqref{eq:Pq_coreshell} in arbitrary units (solid curves),
and the corresponding non-normalized SLS form factors for the considered
temperatures (symbols). The displayed experimental form factors have been obtained by a self-consistent, simultaneous iterative fitting procedure
of the SLS $I(q)$,  and the corresponding DLS data for $D_s(q)$, which is in essence equal to the fitting procedure
used in Ref. \cite{Holmqvist2012}.
The volume fraction is $\phi = 0.14$ for the two main panels of the figure, and $\phi = 0.01$ for the inset in the bottom panel.
}
\end{figure}

The SLS and DLS measurements were performed using a setup by the ALV-Laservertriebsgesellschaft (Langen, Germany).
The static scattered intensity, $I(q)$, and the intensity time autocorrelation function $g_2(q,t)$, 
defined further down in Eq.~\eqref{eq:g2},  
were recorded by an ALV-5000 multitau digital correlator. We carefully checked that there is no noticeable multiple scattering.
In our model which neglects size polydispersity, the mean scattered laser light intensity is described by \cite{BernePecora1976}
\begin{equation}
\label{eq:Iq_static_mono}
   I(q) \propto n\;\!f^2(0)\;\! P(q)\;\! S(q)\,,
\end{equation}
on assuming monodisperse core-shell spheres, with the outer radius, $a_s$, of core plus shell identified with the mean radius $\overline{a}=136$ nm
determined from the x-ray form factor measurement. As pointed out before, to interpret the SLS and DLS data a more detailed form factor
model is needed, different from the one used in our x-ray scattering analysis. 
We use the core-shell form factor model \cite{Pedersen1997}
\begin{equation}\label{eq:Pq_coreshell}
 P(q) = \frac{1}{f^2(0)} \Big\{ \delta_c(T) v(a_c)\;\! b(q a_c) + {\big[} \delta_c(T)-\delta_s(T) {\big ]} \;\! v(a_s)\;\!b(q a_s) \Big\}\,,
\end{equation}
where $f^2(0) = \delta_c(T)\;\! v(a_c) + [ \delta_c(T)-\delta_s(T) ] \;\! v(a_s)$ is the normalization factor,
and $a_c$ and $a_s$ are the radii of core and core plus shell, respectively.
In Eq. \eqref{eq:Pq_coreshell}, $\delta_c(T) = (\epsilon_c(T) - \epsilon_{\text{solv}}(T))/\epsilon_{\text{solv}}(T)$
and $\delta_s(T) = (\epsilon_s(T) - \epsilon_{\text{solv}}(T))/\epsilon_{\text{solv}}(T)$ are the temperature-dependent
core-solvent and shell-solvent relative optical dielectric contrasts,
and $\epsilon_c(T)$, $\epsilon_s(T)$, and $\epsilon_{\text{solv}}(T)$ are optical dielectric permittivities of particle
core, shell and solvent, respectively.
If the temperature is chosen such that $|\epsilon_{\text{solv}}(T) - \epsilon_c(T)|$ is much larger than $|\epsilon_s(T) - \epsilon_c(T)|$,
then the second term in curly brackets is negligible in Eq. \eqref{eq:Pq_coreshell}, and $P(q)$ reduces to the form factor of a homogeneously scattering sphere.

The index-matching of the suspended spheres, required in SLS and DLS, gives rise to a low scattering intensity,
especially for small particle concentrations. Furthermore, direct particle interactions are influential
already at small concentrations, owing to the low suspension salinity and the high effective surface charges of the studied silica particles system.
A SLS measurement of $P(q)$ using  Eq. \eqref{eq:Iq_static_mono}, for conditions where $S(q) \equiv 1$, is thus not feasible.
To determine $P(q)$ in another way, we took advantage of our observation that the wavenumber-dependent
short-time diffusion function determined in the (short-time) DLS measurements is practically temperature independent
in the range between $20-40\;\!^o$C. According to Eq. (\ref{eq:Dsq}), the temperature independence of $D_s(q)/d_0$
implies that also the static structure factor, $S(q)$, contributing to the intensity in Eq. \eqref{eq:Iq_static_mono}
is basically temperature independent in this temperature range. As discussed in Subsec. \ref{subsec:deltagamma},
the $H(q)$ of charged particles is to a good approximation determined by static pair
correlations which in Fourier space are described by $S(q)$. Thus, a temperature independent $S(q)$ implies also a temperature
independent $H(q)$. We have confirmed this implication using our analytic MPB-RMSA calculations of $S(q)$, and our
self-part corrected $\delta\gamma$ method calculations
of $H(q)$ together with the known temperature dependence of the solvent dielectric permittivity
\cite{Akerlof1932, CRChandbook}.
These two theoretical methods for calculating $S(q)$ and $H(q)$, respectively, are quite accurate 
and briefly discussed in Subsecs. \ref{subsec:MPB-RMSA} and \ref{subsec:deltagamma}.

We note that the solvent refractive index, $n_{\text{solv}} \propto \sqrt{\epsilon_{\text{solv}}}$, varies appreciably
in the considered temperature interval, while the refractive indices of the silica core and TPM shell stay practically
constant \cite{Li1994, Wang2012}.
Thus, any variation of the SLS $I(q)$ with temperature is triggered by temperature variations
in $\delta_c(T)$ and $\delta_s(T)$.

As shown in the upper panel of Fig. \ref{fig:SLS_Pq},
the SLS $I(q)$ varies strongly with temperature, owing to the temperature dependence of the
non-normalized form factor $f^2(q)$. In a self-consistent iterative procedure, which in essence is the one
used in \cite{Holmqvist2012}, 
we have simultaneously fitted the SLS intensity $I(q)$, and the DLS data for $D_s(q)/d_0$. The latter DLS data have been fitted using
the self-part corrected $\delta\gamma$ scheme for $H(q)$ in conjunction with the analytic MPB-RMSA input for $S(q)$, and the former
using Eq. \eqref{eq:Iq_static_mono} for $I(q)$ in combination with the MPB-RMSA $S(q)$,
and $P(q)$ according to Eq. \eqref{eq:Pq_coreshell}. The iteration was stopped once
convergence of the iterative results for $D_s(q)/d_0$, $S(q)$, and $P(q)$ had been achieved. The symbols
in the lower panel of Fig. \ref{fig:SLS_Pq} are the results by this iteration procedure for the non-normalized
experimental form factor $f^2(q)$. The curves are the corresponding best fit form factors. The good agreement between experimental
and best-fit form factors in the experimental wavenumber interval, for all considered temperatures,
reflects the consistency of the iterative procedure.
Note that the core-shell structure becomes most visible at $T=30^o$C. Strongest
index matching is obtained for $T \approx 25^o$C where the form factor is seen
to vary only little with increasing $q$. The value for the particle
core radius obtained by the iterative procedure is $a_s= 132$ nm.

The inset in the lower panel of Fig. \ref{fig:SLS_Pq} displays the SLS intensities
measured at the small volume fraction $\phi= 0.01$. Although the particle correlations are rather weak here, 
$I(q)$ is still affected by a non-constant $S(q)$. The scattering intensities at this lower
concentration are overall in good agreement with the self-consistently determined form factors. This confirms additionally
the consistency of our intensity fitting procedure.

In the DLS experiments, we have determined the normalized intensity autocorrelation function,
\begin{equation}
\label{eq:g2}
g_{2} (q, t) = \dfrac{\left<I(q,t_0) I(q,t_0+t) \right>}{{\left<I(q)\right>}^2} \,,
\end{equation}
where $\left< \ldots \right>$ denotes a time average with respect to the starting time $t_0$.
For the fluid-phase ergodic suspensions studied in this work, $g_{2}(q,t)$ decays to zero with increasing
time, and the time average is equivalent to an equilibrium ensemble average.

The  intensity autocorrelation function is related to the normalized electric field autocorrelation function, $g_{1}(q, t)$,
by the Siegert relation \cite{Dhont1996}
\begin{equation}
\label{eq:Siegert}
g_{2} (q, t) = 1 + \tilde{\beta}\;\! g_{1}^2 (q,t)
\end{equation}
where in our experiments the dynamical contrast factor, $\tilde{\beta}$,
is practically equal to one. For a monodisperse suspension
\begin{equation}
\label{eq:Sqt}
g_{1} (q, t) = f_c(q,t) \,,
\end{equation}
i.e. the electric field auto-correlation function is equal to the intermediate scattering function in Eq. (\ref{eq:fct}).

\section{Theory}\label{sec:Theory}

\subsection{Methods used for calculating $\mathbf{S(q)}$}
\label{subsec:MPB-RMSA}

We have calculated the static structure factor, $S(q)$, entering Eq. \eqref{eq:Dsq} for $D_s(q)$ and Eq. \eqref{eq:Iq_static_mono} for $I(q)$,
using the modified penetrating-background corrected rescaled mean spherical approximation (MPB-RMSA) scheme
\cite{Heinen2011}. This semi-analytic and easy-to-implement Ornstein-Zernike integral equation scheme allows for a fast and
accurate calculation of the $S(q)$ of spherical particles interacting by the HSY 
repulsive pair potential in Eqs. (\ref{eq:DLVO}) - (\ref{eq:kappa}), and of the associated radial distribution function $g(r)$. The MPB-RMSA method is convenient for
analyzing experimental scattering curves, and for generating $g(r)$ and $S(q)$ required as input to the analytic-theoretical self-part corrected $\delta\gamma$ and mode-coupling theory schemes discussed further down, and used in our calculation of diffusion properties. 
In Ref. \cite{Heinen2011}, the MPB-RMSA predictions for $S(q)$ were shown to be in excellent agreement
with MC simulation results, and results by the accurate but numerically more expensive Rogers-Young (RY) integral equation scheme, for which convergent
solutions are not always easy to determine \cite{Rogers1984}.

In Subsec. \ref{sec:sub:S_of_q}, we demonstrate the good performance of the MPB-RMSA scheme for the studied  concentration series of silica spheres, by the
comparison with structure factors generated by Monte-Carlo (MC) simulations and the RY scheme, and our SLS results for $S(q)$.

\subsection{Self-part corrected $\mathbf{\delta\gamma}$ method}
\label{subsec:deltagamma}

The influence of HIs on the short-time self-diffusion function $D_s(q)$ determined
by the DLS experiments is encoded in the hydrodynamic function $H(q)$,
which is given in theory by the equilibrium ensemble average \cite{Nagele1996},
\begin{equation}
 H(q) = \lim_\infty \left\langle \frac{1}{N \mu_0} \sum_{l,j=1}^N \hat{\boldsymbol{q}} \cdot
 {\boldsymbol{\mu}_{lj}(\boldsymbol{r}^N)} \cdot \hat{\boldsymbol{q}}
 \exp \left\lbrace i \boldsymbol{q}\cdot[\boldsymbol{r}_l - \boldsymbol{r}_j] \right\rbrace \right\rangle \,,
\label{eq:H_of_q_microscop}
\end{equation}
for a system of $N$ interacting spheres with center positions $\boldsymbol{r}_i$ and $i = 1 \ldots N$ \cite{Dhont1996,Nagele1996}.
Here, ${\boldsymbol{\mu}_{lj}(\boldsymbol{r}^N)}$ is the second-rank mobility tensor
that linearly relates the force on particle $j$ to the translational velocity of particle $i$ \cite{KimKarilla},
and $\mu_0=d_0/(k_BT)$ is the single-sphere translational mobility. The unit vector $\hat{\boldsymbol{q}}$ points 
in the direction of the scattering vector $\boldsymbol{q}$, and  
$\lim_\infty$ denotes the thermodynamic limit $N\to\infty$ and $V\to\infty$,
with $n=N/V$ kept constant, corresponding to macroscopic system volume $V$.
The hydrodynamic function can be written as the sum
\begin{equation}
 H(q) = \frac{d_s}{d_0} + H_d(q) \,,
\label{eq:H_of_q_distinct}
\end{equation}
of a $q$-dependent distinct part $H_d(q)$, and a self-part equal to the reduced short-time self-diffusion coefficient. The function $H(q)$ quantifies the influence of the HIs on short-time colloidal diffusion, as hallmarked by the deviations of $H(q)$ from its infinite dilution value one.
In a non-dilute suspension, $\boldsymbol{\mu}_{lj}$ depends on the instantaneous position super vector, $\boldsymbol{r}^N=\{{\bf r}_1,\cdots,{\bf r}_N\}$,
of the $N$ colloidal particles centers, making it difficult to evaluate the average in Eq. \eqref{eq:H_of_q_microscop} fully analytically. Various semi-analytic theoretical schemes for calculating $H(q)$ are discussed in the literature, based on approximations for
${\boldsymbol{\mu}}_{lj}$ (see, e.g. \cite{Nagele1996, Heinen2011_dyn, Makuch2012, Makuch2015}). 

Our method of calculating $H(q)$ semi-analytically is
an improved, self-part corrected version of the $\delta\gamma$ method by Beenakker and Mazur \cite{Beenakker1983, Mazur1984},
described in detail in \cite{Heinen2011_dyn}. It is a mean-field approach based on the partial resummation of
ring-diagrammatic many-body HIs contributions, and it requires $S(q)$ as its only input. The method was originally applied
to neutral hard spheres only. In subsequent applications to HSY particle systems, 
it was shown to be consistently in good agreement with ASD simulation \cite{Heinen2011_dyn} and experimental results
for $H(q)$ \cite{Westermeier2012}, provided the self-part of $H(q)$ is calculated by using for $d_s/d_0$ a more accurate method than the (zeroth-order) $\delta\gamma$ scheme. This is referred to as self-part
correction. For the low-salinity silica suspensions studied here, $d_s$ has been calculated to good accuracy using the so-called pairwise
additivity (PA) approximation where the two-body HIs contributions to $d_s$ are fully accounted for \cite{Jeffrey1984} but three- and more particle HIs
contributions are disregarded. We refer to \cite{Heinen2011_dyn} for a detailed discussion of the semi-analytic PA method.

Recently, the $\delta\gamma$-scheme has been 
revisited by Makuch and Cichocki \cite{Makuch2012}, who partially lifted the approximations made in Beenakker and Mazur's original derivation of the $\delta\gamma$ scheme. 
However, the results of these more stringent calculations of the $H(q)$ of neutral hard spheres 
only modestly improve those by the original $\delta\gamma$ results, and are obtained at the price of a considerably larger numerical effort.  

\subsection{Collective dynamics}
\label{subsec:Collective dynamics}

A common starting point for the theoretical exploration of the time evolution of the intermediate scattering function $f_c(q,t)$ of
a monodisperse colloidal suspension is the memory equation (see, e.g. \cite{NaegeleMolphys:2002})
\begin{equation}
 \frac{\partial}{\partial\;\!t}\;\!f_c(q,t) = -\;\!q^2\;\!D_s(q)\;\!f_c(q,t) - \int_0^t\!\!du\;\! M_c^{irr}(q,t-u) \;\!
 \frac{\partial}{\partial\;\!u}\;\!f_c(q,u) \,,
\label{eq:memory_eq}
\end{equation}
derived on basis of the generalized Smoluchowski diffusion equation by using the Mori-Zwanzig projection operator formalism.
The first term on the right-hand-side determines the exponential initial decay of $f_c(q,t)$
according to Eq. (\ref{eq:fct-shorttime}), valid in the
colloidal short-time regime $\tau_B\ll t \ll\tau_a$. The integral term on the right-hand-side of Eq. (\ref{eq:memory_eq}) containing the
irreducible collective diffusion memory function, $M_c^{irr}(q,t)$, accounts for memory effects originating from the collective particle motion
across distances $\sim 1/q$. This term gives rise to an overall slower and non-exponential decay of $f_c(q,t)$ at intermediate times
$t \sim \tau_a$, and in general also at long times $t \gg \tau_a$.

The Laplace-transformed memory equation reads
\begin{equation}
  \widetilde{f}_c(q,z) = \frac{1}{z + q^2\;\!D(q,z)} \,,
\label{eq:memory-Laplace}
\end{equation}
with the wavenumber and $z$-dependent diffusion function
\begin{equation}
  D(q,z) = \frac{D_s(q)}{1 + \widetilde{M}_c^{irr}(q,z)} \,.
\label{eq:diffusion-function-Laplace}
\end{equation}
Here, $\widetilde{f}_c(q,z)$ and $\widetilde{M}_c^{irr}(q,z)$ are, respectively,
the temporal Laplace transforms of $f_c(q,t)$ and $M_c^{irr}(q,t)$.

In the fluid colloidal-phase space part, both $f_c(q,t)$ and $M_c^{irr}(q,t)$ decay completely monotonically to zero with increasing time, owing to the overdamped Brownian motion of the colloidal particles and the low-Reynolds number hydrodynamics for which energy dissipation is positive for
arbitrary particle motion \cite{NageleVarenna2013} If, for a given $q$ value, the memory function would
decay sufficiently faster than $f_c(q,t)$, the memory integral in Eq. (\ref{eq:memory_eq}) could be de-convoluted, for $t\gg\tau_a$, according to
\begin{equation}
  \int_0^t\!\!du\;\! M_c^{irr}(q,t-u)\;\!\frac{\partial}{\partial\;\!u}\;\!f_c(q,u)  \approx \widetilde{M}_c^{irr}(q,z=0) \;\!
 \frac{\partial}{\partial\;\!t}\;\!f_c(q,t) \,.
\label{eq:de-convolution}
\end{equation}
This, in turn, would imply a long-time exponential decay, $f_c(q,t\gg\tau_a)\propto \exp\{-q^2\overline{D}(q)t\}$,
of the intermediate scattering function, with the so-called mean collective diffusion function, $\overline{D}(q)$, given by
\begin{equation}
  \overline{D}(q) = D(q,z=0)=\frac{1}{q^2\;\! \overline{\tau}(q)} \,.
\label{eq:Dl_de-convolution}
\end{equation}
The mean relaxation time,
\begin{equation}
  \overline{\tau}(q) = \int_0^\infty\!\! dt\;\!f_c(q,t) \,,
\label{eq:tau-mean}
\end{equation}
characterizes the overall relaxation of $f_c(q,t)$. From Eq. (\ref{eq:diffusion-function-Laplace}), one notices that $\overline{D}(q) \leq D_s(q)$ and hence $\overline{\tau}(q) \geq \tau_s(q)$, with $\tau_s(q)=1/(q^2 D_s(q)$ characterizing the short time relaxation of $f_c(q,t)$.
If de-convolution of the memory integral would be valid at long times, the ratio $\overline{D}(q)/D_s(q)$ would inherit its
$q$-dependence from the $q$-dependence of the time-integrated memory function $\tilde{M}_c^{irr}(q,0)$.

However, since $M_c^{irr}(q,t)$ decays in general not sufficiently faster than $f_c(q,t)$ unless $q$ is very small, the long-time
de-convolution in Eq. (\ref{eq:de-convolution}) is not valid for non-small values of $q$. De-convolution applies rigorously only in the hydrodynamic limit regime of $q \to 0$ and $t \to \infty$ with $q^2\;\!t$ kept constant. 
In this regime, the $f_c(q,t)$ of truly monodisperse spheres decays exponentially according to
$f_c(q,t) = \exp\{-q^2\;\!d_l^c\;\!t\}$, where $d_l^c=\lim_{q \to 0}D(q,z=0)$ is the macroscopic gradient diffusion coefficient. Synthetic colloidal
suspensions are always size polydisperse to a certain extent. In suspensions of strongly correlated particles such as the ones considered here where
the relative osmotic compressibility is small, the decay of the intermediate scattering function at very low $q$ is thus dominated, even for only slightly polydisperse particles, by incoherent scattering contributions related to self-diffusion \cite{Naegele1997}.

The exponential decay of $f_c(q,t)$ in the hydrodynamic regime of  very small wavenumbers $q \ll q_m$ should be distinguished from the long-time
exponential decay predicted in the factorization scaling approximation for non-small wavenumber values around $q_m$ or larger where residual polydispersity has little effect on the intermediate scattering function.
To explore theoretically the validity of factorization scaling, it is advantageous to consider with Cichocki and Felderhof \cite{CichockiFelderhof:1993,CichockiFelderhof:1994} the spectral distribution function (SDF), $p_q(\lambda)\geq 0$,
of relaxation rates $\lambda \geq 0$. The SDF is related to the intermediate scattering function by (see also \cite{Pottier:2011})
\begin{equation}
  f_c(q,t) = \int_0^\infty\!\!d\lambda\;\!p_q(\lambda)\;\! \exp\{-\lambda\;\!t\} \,,
\label{eq:spectral-distribution-function}
\end{equation}
and to the 
Laplace transform of $f_c(q,t)$ by the one-sided Hilbert transform
\begin{equation}
  \widetilde{f}_c(q,z) = \int_0^\infty\!\!d\lambda\;\! \frac{p_q(\lambda)}{z + \lambda}\,.
\label{eq:spectral-distribution-function-Laplace}
\end{equation}
The SDF characterizes the completely monotonic decay of $f_c(q,t)$ characteristic of overdamped Brownian motion.
It is normalized to unity, i.e.
\begin{equation}
  \int_0^\infty\!\!d\lambda\;\! p_q(\lambda) = 1\,,
\label{eq:normality}
\end{equation}
and its first moment is given by 
\begin{equation}
  \big < \lambda \big >_{p_q} = q^2 D_s(q) = q^2 d_0 \frac{H(q)}{S(q)}\equiv \lambda_q^s\,,
\label{eq:first-moment}
\end{equation}
where $H(q)$ is identically equal to one without HIs.
In terms of $\widetilde{f}_c(q,z)$, the SDF reads
\begin{equation}
  p_q(\lambda) = \frac{1}{2\pi i} \left[\widetilde{f}_c(q,-\lambda-i \epsilon) - \widetilde{f}_c(q,-\lambda+i \epsilon) \right]
  = -\frac{1}{\pi} \Im\widetilde{f}_c(q,z=-\lambda+ i \epsilon) \,,
\label{eq:discontinuity}
\end{equation}
where $\epsilon=0^+$ is a positive infinitesimal. The SDF is thus non-zero for relaxation rates $\lambda$ only where $\widetilde{f}_c(q,z)$
is non-analytic at $z=-\lambda$, with its value for these rates determined by the discontinuity of $\widetilde{f}_c(q,z)$ across
the real $z$ axis at $z=-\lambda$.
According to Eqs. (\ref{eq:spectral-distribution-function}) and (\ref{eq:normality}),
the long-time asymptotic decay of $f_c(q,t)$ is determined by the smallest relaxation rate $\lambda$
of non-zero spectral weight $p_q(\lambda)$, or correspondingly by the singularity of $\widetilde{f}_c(q,z)$ in the complex $z$-plane having the largest (real-valued) argument.

In the fluid-phase state, the spectral moment
\begin{equation}
    \int_0^\infty\!\!d\lambda\;\! \frac{p_q(\lambda)}{\lambda} = \widetilde{f}_c(q,z=0)= \overline{\tau}(q) \,,
\label{eq:moment}
\end{equation}
is finite, implying zero spectral weight at $\lambda=0$. This is different from a system at a dynamic glass transition point where $\widetilde{f}_c(q,z)$ has a first-order pole at $z=0$. 

Explicit calculations of $p_q(\lambda)$ were done so far for hard-sphere systems only where HIs are disregarded. 
For smaller $\phi$, Cichocki and Felderhof derived a SDF of the form \cite{CichockiFelderhof:1993}
\begin{equation}
    p_q(\lambda) = a_q \;\!\delta(\lambda - \lambda_q^\ast) + \Delta p_q(\lambda)\;\!\Theta(\lambda -q^2 d_0/2)\,.
\label{eq:SDF-HS}
\end{equation}
It consists of a continuous part, $\Delta p_q(\lambda)$, which is zero in a gap region extending from $\lambda=0$ to $q^2 d_0/2$ as indicated
by the unit step function $\Theta$, and a
$\delta$-function peak contribution of strength $0 \leq a_q < 1$, located at a relaxation rate $\lambda_q^\ast$ inside this gap region.
The non-singular spectral density part, $\Delta p_q(\lambda)$, has for non-zero $\phi$ typically a maximum near the short-time relaxation rate $\lambda_q^s =q^2 D_s(q)$ of $f_c(q,t)$. In the zero concentration limit where the memory function vanishes, $p_q(\lambda) \to \delta(\lambda-\lambda_q^0)$ with $\lambda_q^0=q^2\:\!d_0$.

Eq. (\ref{eq:SDF-HS}) arises since $\widetilde{M}_c^{irr}(q,z)$ is analytic for real-valued $z > -q^2 d_0/2$ located to the right of a square-root type branch point at $z= -q^2 d_0/2$. According to Eqs. (\ref{eq:memory-Laplace}) and (\ref{eq:diffusion-function-Laplace}),
this implies that $\widetilde{f}_c(q,z)$ is analytic for $z > -q^2 d_0/2$, except possibly for an isolated simple pole in this region at
\begin{equation}
    z^\ast = - \lambda_q^\ast = \frac{-q^2 D_s(q)}{1+\widetilde{M}_c^{irr}(q,z^\ast)} \,,
\label{eq:z-ast}
\end{equation}
where the denominator in Eq. (\ref{eq:memory-Laplace}) is zero. Note that $D_s(q)= d_0/S(q)$ without HIs. The pole of $\widetilde{f}_c(q,z)$ at $z^\ast$ gives rise to a  $\delta$ function
contribution to $p_q(\lambda)$ inside the spectral gap region of relaxation rates $\lambda$, provided $S(q)$  and hence $\widetilde{M}_c^{irr}(q,z^\ast)$ at the considered $q$ are large enough for $z^\ast$ to be located inside the region $z > -q^2 d_0/2$ where the Laplace-transformed memory function is analytic. This requires the particles to be sufficiently strongly correlated, i.e. $\phi$ to be large enough, and the associated wavenumber to be selected such that $S(q)$ is sufficiently large.
If these conditions are met,
$f_c(q,t)$ has a genuine long-time exponential decay in the mathematically strict sense, given by $f_c(q,t\gg \tau_a) = a_q \exp\{-q^2 D_l(q) t\}$, with $D_l(q)=\lambda_q^\ast/q^2$ and $a_q$ determined by the coefficient of the simple pole of $\widetilde{f}_c(q,z)$ at $z^\ast$. Note that the complete monotonicity of $f_c(q,t)$ implies that $D_l(q) < \overline{D}(q) < D_s(q)$.
The value of $\Delta p_q(\lambda)$ at the right boundary 
of the spectral gap is then
\begin{equation}
   \Delta p_q(\lambda= q^2 d_0/2+) = \lim_{t\to\infty} \exp\{t\;\!q^2 d_0/2\}\left[ f_c(q,t) - a_q \exp\{-\lambda_q^\ast\;\!t\}\right] \,,
\label{eq:spectral-jump}
\end{equation}
according the initial value theorem of Laplace transforms. For a semi-dilute hard-sphere suspension is $\Delta p_q(\lambda= q^2 d_0/2+)>0$,
i.e. there is a jump discontinuity in the non-singular spectral part which otherwise is continuous.

The occurrence of a long-time exponential mode does not necessarily mean that the other implications of factorization scaling are met. In fact, as shown theoretically for hard-sphere suspensions without HIs in \cite {CichockiFelderhof:1993,CichockiFelderhof:1994}, $D_l(q)/D_s(q)$ reveals pronounced undulations in the $q$-range where $D_l(q)$ exists, and 
these undulations are similar in shape to those of $\overline{D}(q)/D_s(q)$.
If the condition for the existence of a pole of $\widetilde{f}_c(q,z)$ for a $z > -q^2 d_0/2$ is not fulfilled, 
the long-time asymptotic behavior of $f_c(q,t)$ for a semi-dilute hard-sphere system is determined instead by the square-root branch point of $\widetilde{M}_c^{irr}(q,z)$ at $z = -q^2 d_0/2$, implying that $f_c(q,t\gg \tau_a) \propto t^{-3/2} \exp\{- t\;\! q^2 d_0/2\}$. This algebraic-exponential long-time decay is reached, however, only for very long times where the experimental scattering function has decayed already to its noise level.

As discussed in \cite{CichockiFelderhof:1993}, the spectral gap region for the SDF of semi-dilute hard-sphere suspensions is associated physically with the free diffusion of the center of mass (c.o.m.) of two isolated spheres as quantified by the c.o.m. free diffusion coefficient $d_0/2$. To leading order in $\phi$, the gap shows up also for pair potentials different from the hard-core one, provided the HIs are neglected. Free diffusion of the c.o.m. of particle pairs
is not valid any more if HIs are taken into account, or if many-particles direct interaction effects by the environment come into play at larger concentrations.
The non-singular part of the SDF can then assume non-zero values also inside the c.o.m. gap. For concentrated suspensions
of hard spheres without HIs considered, Cichocki and Felderhof \cite{CichockiFelderhof:1994} have proposed the so-called CEA approximation for $f_c(q,t)$,
which captures BD simulation results quite well up to  $\phi =0.4$, in the time window accessed in the simulations. While the $p_q(\lambda)$
in the CEA has the c.o.m. free diffusion gap for all $\phi$, the main features of the spectrum are nonetheless captured correctly also
for moderately large concentrations.

\subsubsection{Mode-coupling theory method}

Different from the CEA method which applies to neutral hard spheres only, the MCT employed in this work can be used for arbitrary pair potentials including the DLVO-type potential of charge-stabilized spheres in Eq. (\ref{eq:DLVO}). Our MCT calculations of collective and self-diffusion properties of suspensions of strongly correlated particles serve two purposes. First they allow to study generic features of diffusion, for genuinely long times $t/\tau_a \gg 10$ not
accessed in the ASD and BD simulations. Second, we can assess the precision of the MCT by the comparison with our BD simulation results.
In the standard MCT for a Brownian suspension where HIs are neglected, the collective memory function in Eq. (\ref{eq:memory_eq}) is approximated by
\begin{equation}
  M_c^{irr}(q,t) = \frac{d_0}{32 \pi^2 n q^3} \int_0^\infty\!\!dk\;\!k\;\!S(k) \int_{|q-k|}^{q+k} dk'\;\!k'\;\!S(k')\;\!\left[V_c(q,k,k') \right]^2
  \;\!f_c(k,t)\;\!f_c(k',t) \,,
\label{eq:MCT-memory-function}
\end{equation}
with static vertex function
\begin{equation}
  V_c(q,k,k') = \left(q^2 +k^2 -{k'}^2 \right) \left[1-\frac{1}{S(k)} \right]  +  \left(q^2 -k^2 +{k'}^2 \right) \left[1-\frac{1}{S(k')} \right] \,,
\label{eq:MCT-vertex}
\end{equation}
isotropic in its three arguments, and symmetric in $k$ and $k'$. The integration variables $k$ and $k'$ in the memory integral are non-negative and cover a strip oriented along the diagonal of the positive $(k'-k)$ plane 
which is bordered at its base by $k'=q-k$, and by $k'=k+q$ and $k'=k-q$ at its upper and lower sides, 
respectively. The base border 
of the strip implies in particular that $k^2 +{k'}^2 \geq q^2/2$. 
The intermediate scattering function in MCT is obtained self-consistently from numerically solving the resulting integro-differential equation. We use here the MPB-RMSA calculated $S(k)$ as the input in Eqs. (\ref{eq:MCT-memory-function}) and (\ref{eq:MCT-vertex}).

While being an approximation, the MCT recovers correctly the weak coupling limit (WCL) of point particles interacting by a Fourier-integrable pair potential $u(r)$, and without HIs considered. The SDF of $f_c(q,t)$ in the WCL has also the c.o.m. free diffusion gap, but different from semi-dilute hard-sphere suspensions there is no delta-function spectral density contribution inside the spectral gap region, for the $\widetilde{M}_c^{irr}(q,z)$ in the WCL is by definition small compared to one. The long-time behavior of $f_c(q,t)$ is here determined by the branch point at $z= - q^2 d_0/2$ which is the singularity of the WCL $\widetilde{M}_c^{irr}(q,z)$ with largest real part. Therefore, $f_c(q,t) \propto \tau_q^{-3/2} \exp\{-\tau_q/2\}$ with $\tau_q=q^2 d_0 t$, and this in turn implies, using Eq. (\ref{eq:spectral-jump}), the continuity of the WCL $p_q(\lambda)$ at the upper gap boundary $\lambda = q^2d_0/2$.

The general structure of $p_q(\lambda)$ can be likewise analyzed using the MCT $\widetilde{M}_c^{irr}(q,z)$ for non-weak coupling conditions. Due to many-particle direct interactions, the $z$-plane gap of $\widetilde{M}_c^{irr}(q,z)$ and hence the SDF gap of zero spectral strength have now a non-trivial upper boundary point different from $q^2d_0/2$ owing to the influence of $S(q)$. Only for $S(q)=1$ is the c.o.m. free diffusion gap width of the SDF recovered. The major contributions to the memory function integral can be expected for $k,\;\!k' \sim q_m$ where the static structure factor is largest. On estimating that $z \lesssim \lambda_q^0/(2\;\!S(q_m))$ along the line of singularities, the expected zero-spectral-strength relaxation gap in $p_q(\lambda)$ shrinks with increasing particle correlations, giving rise to a slower decay of $f_c(q,t)$.  

In our self-consistent calculations of the MCT $f_c(q,t)$ and $f_s(q,t)$, all lengths have been expressed in units of  
the geometric mean particle distance
\begin{eqnarray}
 \overline{r}=n^{-1/3}\,,
\end{eqnarray}
which is the natural length scale for point-like particles, and for the considered HSY spheres 
where the hard-core interaction is masked by the longer-range electric repulsion. 
Different from the WCL, for the strongly correlated particles in this work the MCT-based $\widetilde{M}_c^{irr}(q,z)$ can reveal singularities at $z$ values inside $[-q^2 d_0/2,0]$. 
In fact, for the set of $q$ values considered in Subsec. \ref{sec:sub:Results_longtime} that enclose $q_m$, and for the HSY system parameters considered in this work, the MCT predicts a long-time exponential decay of $f_c(q,t)$, with  $D_l(q)$ determined to good accuracy by $-z^\ast/q^2$ where $z^\ast$ is derived from Eq. (\ref{eq:z-ast}). Within numerical accuracy, the so-obtained $D_l(q)$ agrees with the MCT calculated long-time slope $-(d/dt)\ln S_c(q,t)/q^2$ evaluated numerically using $t \gtrsim 80 \tau_a$.\\ 

For $q \gg q_m$, the memory Eq. (\ref{eq:memory_eq}) for $f_c(q,t)$ reduces to that for the self-intermediate scattering function $f_s(q,t)$ \cite{Nagele1996}. For all wavenumbers, the time evolution of the latter is described by the self-diffusion memory equation for Brownian particles \cite{NaegeleMolphys:2002}
\begin{equation}
 \frac{\partial}{\partial\;\!t}\;\!f_s(q,t) = -\;\!q^2\;\!d_s\;\!f_s(q,t) - \int_0^t\!\!du\;\! M_s^{irr}(q,t-u) \;\!
 \frac{\partial}{\partial\;\!u}\;\!f_s(q,u) \,,
\label{eq:memory_eq_self}
\end{equation}
where $M_s^{irr}(q,t)$ is the non-negative memory function associated with self-diffusion, and $d_s=d_0$ when HIs are neglected. The MSD is related to $f_s(q,t)$ and its memory function by
\begin{equation}
 W(t) = -\lim_{q\to 0}\frac{\log f_s(q,t)}{q^2}\,.
\label{eq:MSD_from_fs}
\end{equation}
The long-time slope of $W(t)$ can be directly expressed in terms of the memory function as 
\begin{equation}
  d_l = \frac{d_s}{1 + \int_0^\infty\!\!dt\;\!M_s^{irr}(q\to 0,t)}\,,
\label{eq:dl_memory}
\end{equation}
showing that $d_l < d_s$ for interacting particles.
 
In standard MCT without HIs used in this work, the self-diffusion memory function is given by \cite{NaegeleDhont:1998}
\begin{equation}
   M_s^{irr}(q,t) = \frac{d_0}{4 \pi^2 n} \int_0^\infty dk\;\!k^4\; S(k)\left(1 - \frac{1}{S(k)} \right)^2 
   \int_{-1}^{1} d\nu\;\!\nu^2 f_c(k,t) f_s\left(\sqrt{q^2+k^2-2qk\nu},t\right)\,,
\label{eq:MCT-memory-function-self2}
\end{equation}
invoking $f_c(q,t)$ in the integral kernel. The coupled MCT integro-differential equations are numerically solved for $f_c(q,t)$ and $f_s(q,t)$, using the MPB-RMSA $S(k)$ as input.

\section{Employed simulation methods}\label{sec:Simulations}

\subsection{Monte-Carlo and Brownian Dynamics simulations}\label{sec:sub:MCBD}

The MC calculations of $S(q)$ were performed for varying 
numbers $N=125 - 512$ of particles placed in a periodically replicated cubic simulation box of reduced length $L/a=[4\;\!\pi\;\!N/(3\;\!\phi)]^{1/3}$, 
in order to minimize residual finite system size effects. The MC generated particle configurations representing equilibrated systems where
used in addition as input to the ASD simulation calculations of the hydrodynamic function $H(q)$ and its large-$q$ asymptotic limit $d_s$. \\

The BD simulations were performed using the ASD simulation method described below, for the special case of diagonal hydrodynamic friction matrices 
describing spheres without HIs. The BD calculations of the various diffusion properties discussed in this paper were all performed using $N=512$ particles in a cubic primary 
simulation box. 

\subsection{Accelerated Stokesian dynamics simulations with finite-size scaling}\label{sec:sub:ASD}

We employ here the accelerated Stokesian Dynamics (ASD) simulation method by Sierou and Brady \cite{SierouBrady2001}, extended to Brownian particles systems by Banchio and Brady \cite{BanchioBrady2003}. 
As we discuss in the following, the ASD simulations of the diffusion coefficients and intermediate and self-intermediate scattering functions are  significantly
dependent on the system size, i.e. the  number of particles, $N$, in the primary cubic simulation box of volume $V=L^3$ that is periodically
replicated in space. The deviations of the properties at finite $N$ from the macroscopic system (i.e. hydrodynamic limit)
values are due to the spatially correlated flow fields of the periodic images caused by the long-range HIs. These deviations are
of order $a/L\propto(\phi/N)^{1/3}$. Regarding the hydrodynamic function, the finite-size correction formula
\begin{equation}
\label{eq:fs-corr_H_q}
H(q) = H_N(q) + 1.76 \frac{\eta_0}{\eta_\infty} S(q) \left(\phi_N\right) ^{1/3} + {\cal O}(\phi_N) \,,
\end{equation}
is well established for a variety of systems \cite{Ladd:1990,LaddWeitz:1995,Banchio2008,Abade2010,Abade2011,Wang2015}. Here, $\phi_N=\phi/N$ has been introduced for brevity. 
For $q \gg q_m$, the correction formula for $H(q)$ reduces to \cite{Ladd:1990}
\begin{equation}
\label{eq:fs-corr_d_s}
\frac{d_s}{d_0} = \frac{d_{s,N}}{d_0} + 1.76 \frac{\eta_0}{\eta_\infty(\phi)} \left(\phi_N\right) ^{1/3} + {\cal O}(\phi_N) \,,
\end{equation}
where $H_N(q)$ and $d_{s,N}$ are the hydrodynamic
function and short-time self-diffusion coefficient, respectively, as obtained from the ASD simulation using $N$ spheres,
and $H(q)$ and $d_s$ are the searched for respective infinite system quantities. 
The subindex $N$ labels quantities obtained in the ASD simulation for $N$ particles in the simulation box.
If the simulated $H_N(q)$ for different  
particle numbers $N$ are corrected by the ${\cal O}((\phi/N)^{1/3})$ term on the right-hand side of Eq. (\ref{eq:fs-corr_H_q}),
the resulting curves collapse practically on a single master curve which can be identified as $H(q)$.
\begin{figure}
\includegraphics[width=0.45\columnwidth,angle=0]{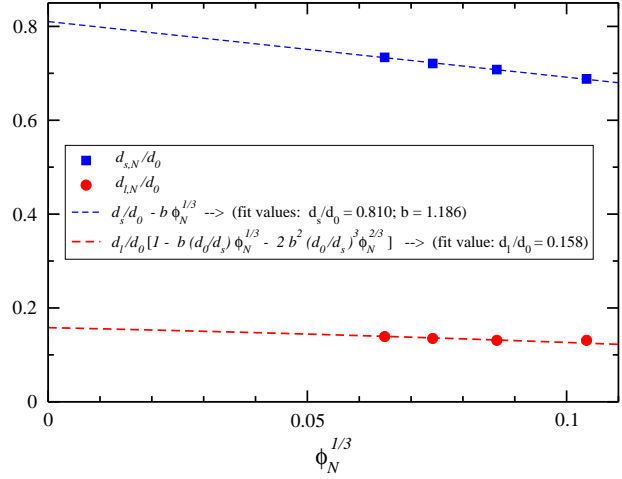}  
\caption{\label{fig:ASD_dl_scaling}
System-size dependence of the ASD simulated long-time and short-time self-diffusion coefficients 
$d_{l,N}$ and $d_{s,N}$ as functions of $\phi_N^{1/3}$ where $\phi_N=\phi/N$, for 
$\phi=0.14$ and numbers $N=\{125,216,343,512\}$ of particles in the primary simulation box.
The lower dashed curve is the fit to the ASD data for $d_l$
(circles) according to Eq. (\ref{eq:dl-correction}). The upper dashed curve is the fit to the ASD data for $d_s$ (squares)
according to Eq. (\ref{eq:fs-corr_d_s}). Thickness of symbols is comparable to numerical uncertainties in the simulation data. }
\end{figure}

The high-frequency suspension viscosity, $\eta_\infty(\phi)$, entering the finite-size correction expression in units of
the suspending fluid shear viscosity $\eta_0$, accounts for the fact that the HIs between a sphere and its periodic images
are at distances of order $L \gg a$  shielded by other particles in the suspension. Different from the hydrodynamic
function, $\eta_\infty(\phi)$ is insensitive to the system size so that, if known, it can be used as input to the correction
formula. Otherwise, $\eta_\infty$ can be estimated from the slope of a straight line fitting of $d_{s,N}$ as a function of
$N^{-1/3}$, as done in \cite{Abade2010}.

Similar to the short-time quantity $H(q)$, we observe a significant $N$-dependence of the ASD simulations results also for the
(self)-intermediate scattering function and the MSD. As a generalized finite-size correction formula for these properties, we use
\begin{eqnarray}
w_c(q,t) = w_{c,N}(q,t) \frac{D_s(q)}{D_{s,N}(q)} =  w_{c,N}(q,t)\frac{H(q)}{H_N(q)}  \label{eq:fs-corr_wc} \\
\end{eqnarray}
which includes the finite-size correction formulas
\begin{eqnarray}
w_s(q,t) &=& w_{s,N}(q,t) \frac{d_s}{d_{s,N}} \label{eq:fs-corr_ws}   \\
W(t) &=& W_N(t)  \frac{d_s}{d_{s,N}}  \label{eq:fs-corr_W}
\end{eqnarray}
for the self-diffusion properties $w_s(q,t)$ and $W(t)$ as limiting cases.
The macroscopic system results for $f_c(q,t)$, $f_s(q,t)$ and $d(t)=\dot{W}(t)$ are directly obtained from the
system-size-corrected simulation results for $w_c(q,t)$, $w_s(q,t)$ and $W(t)$, respectively.

For $t>0$, Eq. (\ref{eq:fs-corr_wc}) expresses in particular the
time independence of the ratio $w_c(q,t)/w_{c,N}(q,t)$, empirically found in our ASD simulations up to statistical fluctuations.
Eq. (\ref{eq:fs-corr_wc}) implies furthermore that the leading-order system size corrections to $w_c(q,t)$, $w_s(q,t)$ and
$W(t)$ are all of ${\cal O}([\phi/N]^{1/3})$. We demonstrate this for the long-time self-diffusion coefficient. Starting
from $d_l/d_{l,N}=d_s/d_{s,N}$ as  implied by Eq. (\ref{eq:fs-corr_W}), a finite-size
correction expression for $d_l$ is obtained given by
\begin{eqnarray}
\label{eq:dl-correction}
 \frac{d_l}{d_0} = \frac{d_{l,N}}{d_0} + \left(\frac{d_l}{d_s}\right)\;\!b\;\!\phi_N^{1/3} + 2\;\!\frac{d_l}{d_s}\left(\frac{d_0}{d_s}\;\!b\right)^2\;\!\phi_N^{2/3} + {\cal O}(\phi_N) \,,
\end{eqnarray}
where $b=1.76\times(\eta_0/\eta_\infty)$.
Thus, the leading-order correction term of $d_{l,N}$ is by the factor $d_l/d_s$ smaller than the leading-order
finite-size correction of $d_{s,N}$. The second-order correction term of $d_l$ proportional to $\phi_N^{2/3}$ is included in order to maintain,
for consistency, all terms up to linear order in $\phi_N$. 

To illustrate the employed finite-size scaling, the system-size dependencies of $d_{l,N}$ and $d_{s,N}$ are shown in
Fig. \ref{fig:ASD_dl_scaling}, for a HSY system with $\phi=0.14$, $Z=175$, and $\kappa a=2.72$ (see Table \ref{tab:sim_param}), and using the ASD simulation results for four different values of $N$ as indicated in the figure. The fit of the
simulation points for $d_{s,N}$, using Eq. (\ref{eq:fs-corr_d_s}) with $d_s$ and $\eta_\infty/\eta_0$ treated as the fit parameters,
leads to a value for $d_s$ listed in Table \ref{tab:sim_tcoef} for the considered volume fraction. The value for $\eta_\infty$ obtained
in this fitting procedure is close to that obtained by our direct ASD viscosity calculation, with the latter listed in the table. We have
used the results for $d_s$ and $\eta_0/\eta_\infty$ obtained in the finite-size correction procedure for $d_s$ as input
to Eq. (\ref{eq:dl-correction}), with $d_l$ treated now as the only fit parameter. This results in the dashed lower straight line in
Fig. \ref{fig:ASD_dl_scaling}, with the resulting value for $d_l$ given in Table \ref{tab:sim_tcoef}.
\begin{figure}[t]
\includegraphics[width=0.45\columnwidth]{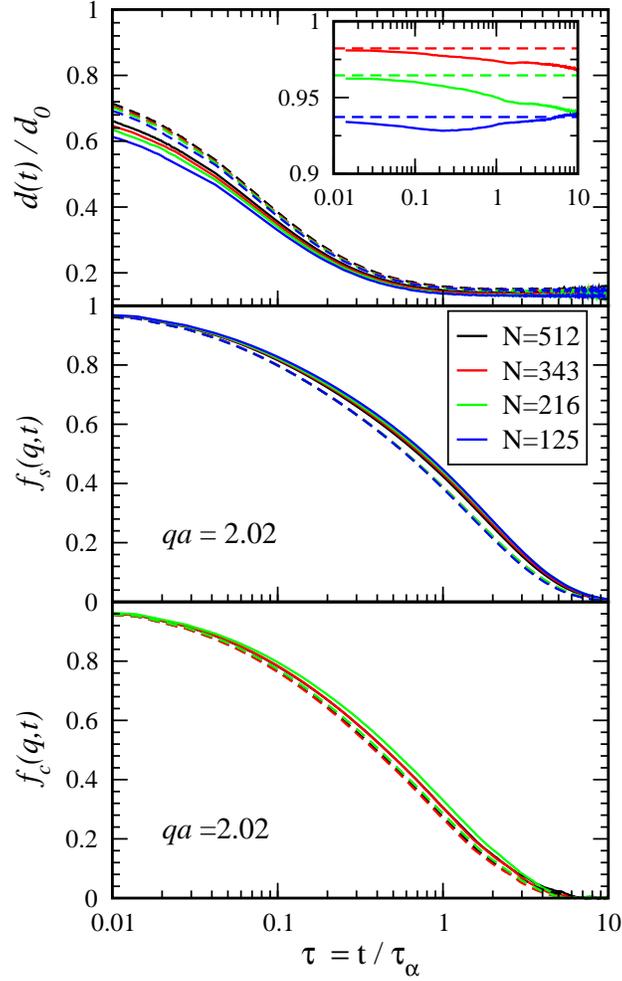}
\caption{ \label{fig:ASD_fs_scaling}
Test of the finite-size scaling expressions in Eqs. (\ref{eq:fs-corr_wc}) - ({\ref{eq:fs-corr_W}}),
using ASD simulation data (solid curve) for $W(t)$ (inset in top panel), its derivative $d(t)$ (top panel), self-intermediate
 scattering function $f_s(q,t)$ (middle panel) and intermediate scattering function $f_c(q,t)$ (bottom panel). Four different particle numbers, $N$, per simulation box are considered and distinguished by different colors. 
The silica system
with $\phi=0.14$, $Z=175$ and $\kappa a =2.72$ is considered, for $qa=2.02$ in the lower two panels 
corresponding to $q/q_m=0.9$. Dashed curves are the finite-size corrected ASD results, as functions of correlation time $\tau=t/\tau_a$ expressed in units of $\tau_a$.
Inset in top panel: $W_{N_1}(t)/W_{N_2}(t)$ for $N_1=\{125,216,343\}$ and $N_2=512$ (solid curves) compared against $d_{s,N_1}/d_{s,N_2}$ (dashed horizontal lines).}
\end{figure}

In Fig. \ref{fig:ASD_fs_scaling}, ASD simulation results for $d_N(t)=\dot{W}_N(t)$, $f_{s,N}(q,t)$ and $f_{c,N}(q,t)$ are shown for four different particle numbers $N$ as indicated (solid curves), 
as functions of the reduced correlation time $\tau=t/\tau_a$.
The finite-size corrected functions obtained by means of Eqs. (\ref{eq:fs-corr_wc})-({\ref{eq:fs-corr_W}}) are depicted as colored
dashed curves, with a different color for each $N$. The dashed 
curves practically coincide except those for $f_c(q,t)$ where some
residual spreading is observed (bottom panel). We attribute this spreading to the, in comparison with $f_s(q,t)$, larger statistical uncertainties for the collective
property $f_c(q,t)$. The statistics in our ASD simulation data for $W(t)$ is good enough that we can numerically
determine its time derivative $d(t)=\dot{W}(t)$, referred to as the self-diffusion function, which is shown in the top panel.
From general properties of the generalized Smoluchowski equation describing the overdamped dynamics of Brownian spheres,
it follows that \cite{CichockiHinsen:1992}
\begin{equation}
\label{eq:self-diffusion-function}
 \frac{d(t)}{d_s} = \frac{\dot{W}(t)}{d_s} = d_\ast + \left(1-d_\ast\right) \gamma(t) \,,
\end{equation}
where $d_\ast=d_l/d_s$. Here, $\gamma(t)$ is a function whose derivative is proportional to the regular part of the particle velocity autocorrelation function and which decays completely monotonically from $\gamma(0)=1$
towards $\gamma(\infty)=0$. The self-diffusion function $d(t)$ decays from its initial value $d_s$ towards its long-time asymptotic
value $d_l$ faster than $W(t)/t$ by a factor of order $1/\sqrt{t}$, with both functions sharing the same initial and long-time values.
And indeed, according to the top panel of Fig. \ref{fig:ASD_fs_scaling},  $d(t)$ approaches its long-time asymptotic
value practically already for $\tau \sim 3$. The inset in this panel shows the MSD ratio $W_{N_1}(t)/W_{N_2}(t)$ for $N_1=\{125,216,343\}$, and $N_2=512$ kept constant (solid curves), 
in comparison with the corresponding short-time self-diffusion coefficient
ratio $d_{s,N_1}/d_{s,N_2}$ (dashed horizontal curves). The very weak time dependence of the three depicted MSD ratios exemplifies the consistency of the simulation data 
with the finite-size correction expression in Eq.~(\ref{eq:fs-corr_W}). 

We emphasize that the ASD simulation data shown in this paper are the results of very time-consuming, elaborate calculations.
Simulations of long-time collective properties typically took of the order of two months on a desktop PC, with the code parallelized
for using 4 CPU cores.

\begin{table}
%{
\footnotesize{ % Just for OneColumn
\caption{System parameters used in our simulations and theoretical calculations:
Particle volume fraction $\phi$, modulus of effective particle charge number, $Z$, and reduced
screening parameter, $\kappa a$, calculated using Eq. (\ref{eq:kappa}). For the considered systems, $T=20^o$C, $a=136$ nm and a fixed value $n_s = 0.7 \mu$M for the
1-1 electrolyte concentration are used. 
The Bjerrum length of the toluene:ethanol mixture is $L_B = 8.64$ nm. For the comparison with DLS data, the experimental value $d_0=2.54\;\!\mu\text{m}^2/\text{s}$ for the single-particle diffusion coefficient is used corresponding to $\tau_a=a^2/d_0=7.28$ ms.\\
} \label{tab:sim_param}}
\vspace{.2em}
\centering
{\footnotesize{
\begin{tabular}{@{\extracolsep{\fill}}lll}
\hline
$\boldsymbol{\phi}$\qquad~~~~&
$\boldsymbol{Z}$\qquad~~~&
$\boldsymbol{\kappa a}$\\
\hline\hline
0.057  &  135  & 1.83  \\
0.14   &  175  & 2.72  \\
0.15   &  190  & 2.90  \\
\hline
\end{tabular}
}}
\end{table}

\begin{table}[!h]
%{
\footnotesize{ % Just for OneColumn
\caption{Finite-size corrected ASD simulation results for the long-time and short-time self-diffusion coefficients $d_l$ and $d_s$, divided by the Stokes-Einstein single-sphere diffusion coefficient $d_0$, 
and for the high-frequency suspension viscosity $\eta_\infty$, divided by the solvent viscosity $\eta_0$. The listed MCT and BD results are without HIs so that $d_s=d_0$ and $\eta_\infty=\eta_0(1+2.5\;\!\phi)$. 
Notice the enhancement of $d_l$ caused by the HIs.
} \label{tab:sim_tcoef}}
\vspace{.2em}
\centering
{\footnotesize{
\begin{tabular}{@{\extracolsep{\fill}}llllll}
\hline
\multirow{2}{*}{$\boldsymbol{\phi}$\qquad~~}&
\multicolumn{3}{c}{ASD\qquad~~}&~BD\qquad~&MCT\\
& $\mathrm{\eta_{\infty}/\eta_0}$\qquad~ &
$\mathrm{d_s/d_0}$\qquad~ &
$\mathrm{d_l/d_0}$\qquad~~&
$\mathrm{d_l/d_0}$\qquad~~&
$\mathrm{d_l/d_0}$\\
\hline\hline
0.057 & 1.15 & 0.93 & 0.48 & 0.40 & 0.40 \\
0.14 & 1.42 & 0.81 & 0.16 & 0.13 & 0.07 \\
0.15 & 1.46 & 0.78 & 0.11 & 0.10 & 0.04\\
\hline
\end{tabular}
}}
\end{table}

\section{Results}\label{sec:Results}

\subsection{Static structure factor}\label{sec:sub:S_of_q}
\begin{figure}
\centering
\includegraphics[width=0.45\columnwidth]{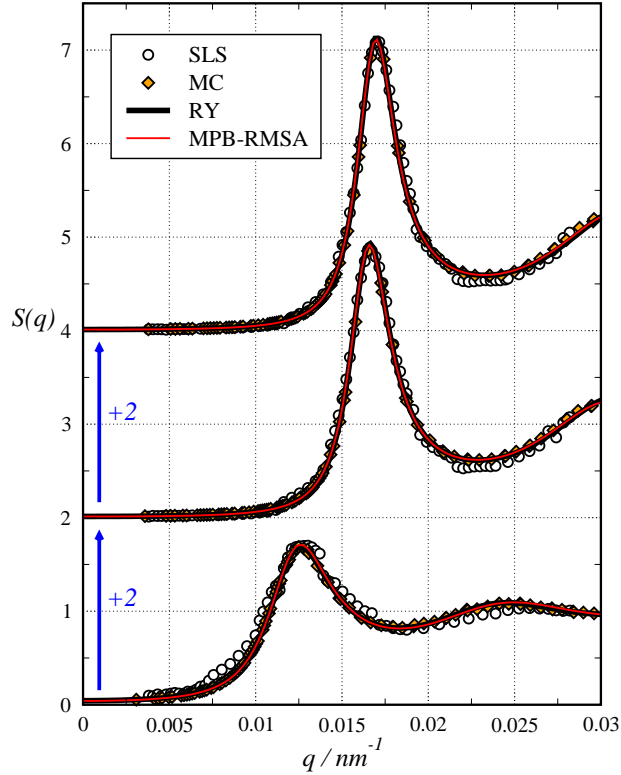}
\vspace{0.0em}
\caption{\label{fig:Sq}
Open circles: Experimental SLS static structure factors of suspensions of silica particles at volume fractions
$\phi = 0.057, 0.14$ and $0.15$ (from bottom to top), and $T=20^o$C.
Black diamonds filled in orange: corresponding Monte Carlo (MC) simulation results.
Black and red solid curves: Rogers-Young (RY) and MPB-RMSA static structure factors, respectively, for the same interaction
parameters used in the MC calculations and given in Table \ref{tab:sim_param}.
For better visibility, the upper two curves are up-shifted by $2$ and $4$, respectively.
The data for $\phi =0.14$ and $0.15$ are partially reproduced from \cite{Heinen2011,Holmqvist2010}.
}
\end{figure}

In Fig. \ref{fig:Sq}, the SLS static structure factors, $S(q)$, of the three silica-spheres suspensions with volume fractions
$\phi = 0.057, 0.14$ and $0.15$ are represented by open circles. The same figure features the corresponding
MC simulation results (black diamonds filled in orange), RY scheme \cite{Rogers1984} results (black solid curves) and
MPB-RMSA-scheme results \cite{Heinen2011} (thin solid red curves). 
For each $\phi$, identical HSY pair-potential parameters $\{\kappa, a, Z, L_B\}$ were used in the calculation of the MC, RY, and MPB-RMSA results displayed in Fig. \ref{fig:Sq},
giving rise to structure factors that are nearly indistinguishable on the scale of the figure, 
and that likewise are in excellent overall agreement with the SLS data. The system parameters used in our simulations and theoretical calculations are listed in Table \ref{tab:sim_param}. 

The principal peak wavenumber positions, $q_m$, of the static structure factors displayed in Fig. \ref{fig:Sq}, are in
agreement with the limiting $\phi$-scaling expression $q_m\sigma \lesssim 2.2 \times \pi^{2/3} \times {(6\phi)}^{1/3}$ for charged
spheres at low salinity \cite{Heinen2011, Westermeier2012}. The principal peak position, $q_m$, for each of the three  SLS $S(q)$'s obeys the
upper limit set by the above scaling expression proportional to $\phi^{1/3}$, with a maximal undershoot of $4\%$ observed for the least concentrated system with $\phi = 0.057$.
This indicates that long-ranged electrostatic repulsion is the dominant inter-particle force in each of the three silica suspensions studied here. The system at $\phi=0.15$ is close to the fcc freezing transition taking place at $\phi\approx 0.16$. This is indicated by the associated principal peak value $S(q_m)\approx 3.1$ given in Table \ref{tab:fitvalues}, in accord with the empirical Hansen-Verlet peak value freezing criterion for low-salinity HSY suspensions \cite{Gapinski2014,NaegeleMolphys:2002}. The strong electric inter-particle repulsion causes the relative osmotic compressibility, $S(q\to 0)$, to be very small as it is noticed in the figure.\\ 
 
\subsection{Short-time collective diffusion}\label{sec:sub:Results_shorttime}

\begin{table}
%{
\footnotesize{ % Just for OneColumn
\caption{Experimental volume fraction, $\phi$, and static structure factor maximum, $S(q_m)$, and experimentally extracted hydrodynamic function principal peak height, $H(q_m)$, reduced short-time diffusion function minimum, $D(q_m)/d_0$, and reduced short-time self-diffusion coefficient, $d_s/d_0$, for the considered concentration series of silica-spheres suspensions. Additionally shown is the self-part corrected $\delta\gamma$ method prediction for the sedimentation coefficient $H(0)=H(q\to 0)$.\\
}\label{tab:fitvalues}}
\vspace{.2em}
\centering
{\footnotesize{
\begin{tabular}{@{\extracolsep{\fill}}llllll}
\hline
$\phi$\qquad~~~~&
%$\mathrm{\mathbf{n_s} [\mu}M]$\qquad~~~~&
$\mathrm{S(q_{m})}$\qquad~&
$\mathrm{H(q_{m})}$\qquad~&
$\mathrm{D_s(q_{m})/d_0}$\qquad~&
$\mathrm{d_s/d_0}$\qquad~&
$\mathrm{H(0)}$\quad\\
\hline\hline
0.057  & 1.71  &  1.16  &  0.87  & 0.93 & 0.43\\
0.14   & 2.90  &  1.16  &  0.38  & 0.77 & 0.27\\
0.15   & 3.11  &  1.15  &  0.36  & 0.75 & 0.25\\
\hline
\end{tabular}
}}
\end{table}

\begin{figure}
\centering
\includegraphics[width=0.45\columnwidth]{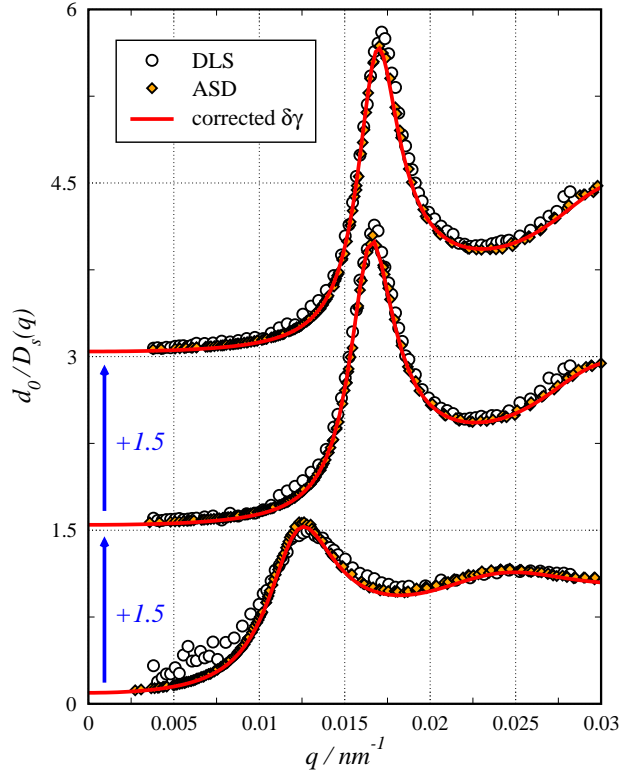}
\vspace{-1em}
\caption{\label{fig:Dq}
Open circles: Experimental inverse reduced short-time diffusion function, $d_0 / D_s(q)$, obtained using DLS 
applied to silica-spheres suspensions at $\phi = 0.057, 0.14$ and $0.15$ (from bottom to top). 
Black diamonds filled in orange: ASD simulation results.
Red solid curves: Results by the self-part corrected $\delta\gamma$-scheme, with $d_s/d_0$ calculated using the PA-scheme. As the only input to this scheme, the MPB-RMSA $S(q)$'s of Fig. \ref{fig:Sq} were used.
For better visibility, the upper two curves are up-shifted by $1.5$ and $3$, respectively.
The depicted data for $\phi = 0.14$ and $0.15$ are partially reproduced from \cite{Holmqvist2010} and \cite{Heinen2010}.
}
\end{figure}
\begin{figure}
\centering
\includegraphics[width=0.45\columnwidth]{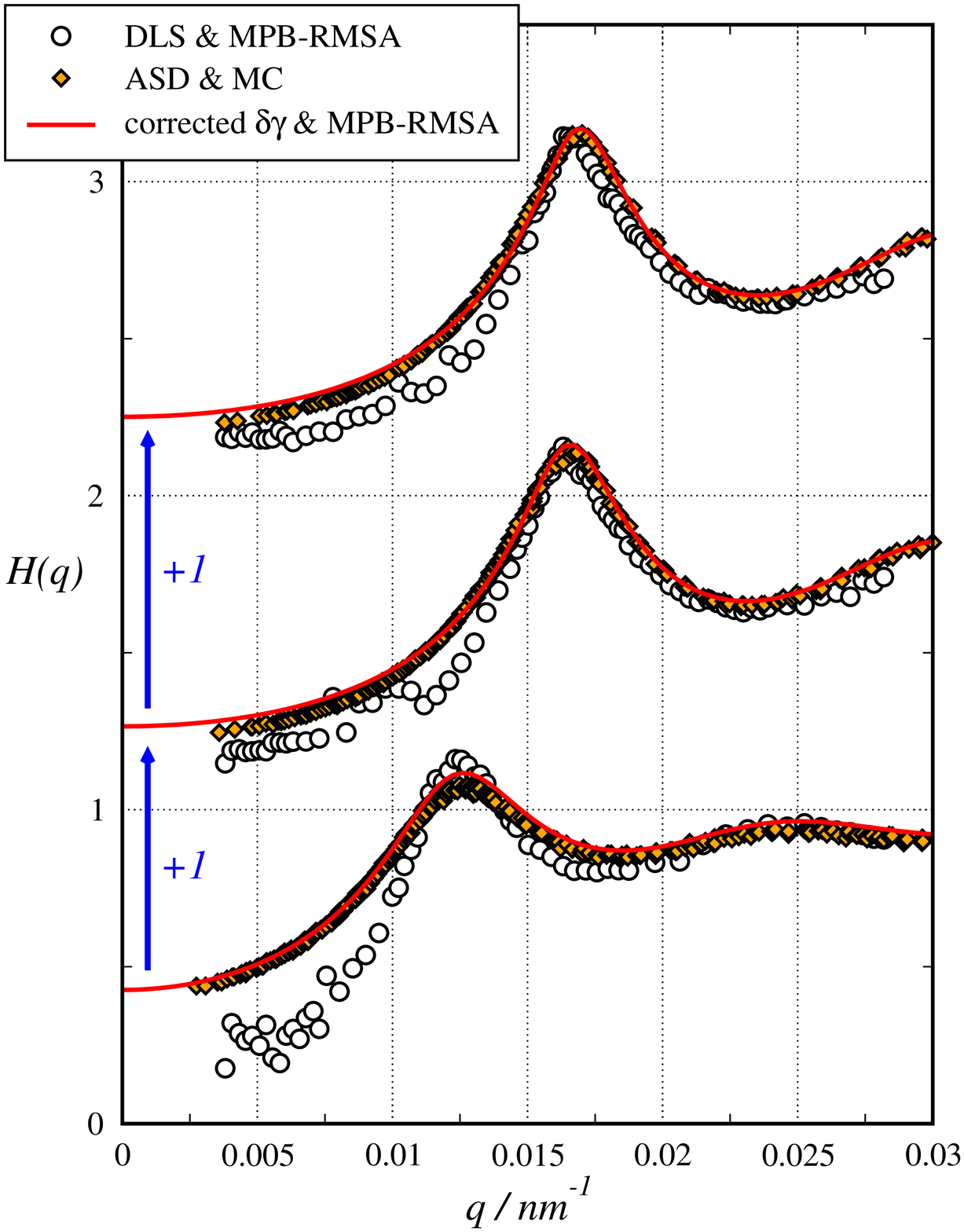}
\vspace{-1em}
\caption{\label{fig:Hq}
Open circles: Hydrodynamic function, $H(q)$, obtained from dividing the MPB-RMSA $S(q)$'s in Fig. \ref{fig:Sq} 
by the
DLS data for $d_0 / D_s(q)$ in Fig. \ref{fig:Dq}, for $\phi = 0.057, 0.14$ and $0.15$ (from bottom to top). 
Black diamonds filled in orange: ASD simulation results.
Red solid curves: Self-part corrected $\delta\gamma$ scheme results with MPB-RMSA $S(q)$ input,
and with reduced short-time self-diffusion coefficient, $d_s/d_0$, according to the PA-scheme.
For better visibility, the upper two curves are up-shifted by $1$ and $2$, respectively.
}
\end{figure}
The DLS data for the inverted reduced short-time diffusion function, $d_0/D_s(q)$ at the considered three volume fractions are depicted in Fig. \ref{fig:Dq}, and compared against corresponding ASD simulation and self-part corrected $\delta\gamma$ method predictions with MPB-RMSA $S(q)$ input. There is excellent agreement between simulation and theoretical results, and they in turn agree well with the DLS data, except for the low-$q$ data of the least concentrated system for which the statistical fluctuations are largest. The good overall agreement, regarding $D_s(q)$, between theory/simulation and experiment is remarkable, considering that no adjustable parameter is invoked in the comparison. Without HIs, $d_0/D_s(q)=S(q)$, so that the differences in the respective curves in Figs. \ref{fig:Dq} and \ref{fig:Sq} are due to hydrodynamics. The principal peak of $S(q)$ corresponds to a principal minimum of $D_s(q)$ at practically the same wavenumber $q_m$. The latter is referred to as the cage diffusion coefficient, $D_s(q_m)$, since it is related to the width $\sim 2\pi/q_m$ of the nearest neighbor cages of particles formed around each particle. The experimental values of $D_s(q_m)$ are listed in Table \ref{tab:fitvalues}, describing the decrease of $D_s(q_m)$ with increasing $\phi$ as the cage becomes stiffer (i.e., less mobile). As discussed in \cite{Gapinski2010}, the fluid-crystal freezing point in low-salinity charge-stabilized systems is reached when the cage diffusion coefficient falls below the critical value $D_s(q_m)/d_0 \approx 1/3$. While we are dealing here with low-salinity HSY systems only, where  particle correlations are strongest, it is instructive to discuss the effect of adding salt: with increasing concentration of added salt, i.e. decreasing electrostatic screening length $1/\kappa$, $D_s(q_m)$ decreases monotonically for given $\phi$ until at very strong electric screening where $\kappa a \gg 1$ the hard-sphere limiting value $D_s^\text{HS}(q_m)$ of the cage diffusion coefficient is approached for which an accurate analytic expression is known  \cite{Banchio1999,NaegeleMolphys:2002} applying to the full fluid phase concentration range. For $\phi \leq 0.3$, the hard-sphere cage coefficient decreases linearly with increasing concentration according to $D_s^\text{HS}(q_m)\approx 1 - 2\;\!\phi$. At larger $\phi$, it decreases less than linearly until at the freezing volume fraction $\phi_f \approx 0.494$ the hard-sphere freezing value $D_s^\text{HS}(q_m) = 0.117$ is reached.

In Fig. \ref{fig:Hq} the hydrodynamic functions $H(q)=\left(D_s(q)/d_0\right)\times S(q)$ of the concentration series are depicted, obtained from the $D_s(q)$'s of the previous figure by multiplication with the corresponding $S(q)$ as indicated in the figure caption. There are significant undulations of $H(q)$ caused by the HIs. Noteworthy here are the peak values $H(q_m)$ in Table \ref{tab:fitvalues} larger than one that are characteristic for low-salinity systems at small to moderately large volume fractions \cite{Gapinski2010, Westermeier2012}. This should be contrasted with the HSY high-screening limiting case of hard spheres for which $H^\text{HS}(q_m)= 1 - 1.35\;\!\phi$ is valid to excellent accuracy up to the freezing volume fraction \cite{Banchio1999}. While the general ordering relations $H(q_m) > H^\text{HS}(q_m)$ and $d_s > d_s^\text{HS}$ hold for the hydrodynamic function peak height and self-diffusion coefficient, respectively, the opposite ordering $H(0) < H^\text{HS}(0)$ is valid for the (short-time) sedimentation coefficient $H(0)=H(q\to 0)$, expressing that in a homogeneous suspension of slowly settling charged particles the mean settling velocity is smaller than in a corresponding suspension of electrically neutral 
particles. Theoretical estimates of $H(0)$ gained using the corrected $\delta\gamma$ scheme are listed in Table \ref{tab:fitvalues}. They describe the decrease of the sedimentation coefficient with increasing concentration.   

Table \ref{tab:fitvalues} includes also the normalized short-time self diffusion coefficient, $d_s/d_0$, extracted from our SLS/DLS experiment results.
Due to the limited $q$-range accessible in light scattering, $d_s/d_0 = \lim_{q\to\infty}D_s(q)/d_0$
can be extracted from the experimental DLS data only indirectly. This is clearly seen in Fig. \ref{fig:Dq}, 
where $d_0/D_s(q)$ features pronounced undulations around its large-$q$ asymptotic value $d_0/d_s$, visible even near  to the largest experimentally accessed 
wavenumber $q \approx 0.0275$ nm$^{-1}$.
A simple method to extract $d_s/d_0$ approximately from the knowledge of $D_s(q)$ in a limited $q$-range has been introduced by Pusey \textit{et al.} in \cite{Pusey1978, Segre1995}, and further rationalized theoretically in \cite{Abade2010}. The method has been used for charge-stabilized  particles, e.g., in \cite{Heinen2010,Holmqvist2010}.  In Pusey's method, $d_s$ is identified as $D_s(q^*)$, where $q^*$ is the wavenumber at which
the right shoulder of the principal peak in $S(q)$ passes through unity (\textit{c.f.}, Fig. \ref{fig:Sq}, with $q^* \approx 0.015$ nm$^{-1}$ for the bottom curve, and
$q^* \approx 0.02$ nm$^{-1}$ for the upper two curves).
With the $\delta\gamma$ scheme at our disposal, in the present work we use a convenient alternative way to determine $d_s/d_0$ that was used 
already in Ref. \cite{Genz1991}, and shown recently \cite{Westermeier2012} to be quite accurate for charged-sphere suspensions in the entire fluid phase regime.
It consists of treating $d_s/d_0$ as an adjustable parameter in fitting ${H_d(q)|}_{\delta\gamma} + d_s/d_0$ to the
hydrodynamic function $H(q)$ determined from the DLS-recorded $D_s(q)/d_0$, and the static structure factor $S(q)$. Here, ${H_d(q)|}_{\delta\gamma}$ is the wavenumber-dependent distinct part of the hydrodynamic function evaluated using the $\delta\gamma$ scheme with MPB-RMSA input for $S(q)$. The so-determined values for $d_s/d_0$ listed in Table  \ref{tab:fitvalues} compare well with the ASD predictions for $d_s/d_0$ in Table \ref{tab:sim_tcoef}. Owing to HIs, $d_s$ increasingly falls below its infinite dilution value $d_0$ with increasing concentration.   

\subsection{Collective diffusion at intermediate and long times}\label{sec:sub:Results_longtime}

In the following, ASD and BD simulation, MCT and experimental results are discussed for the silica
particles system of volume fraction $\phi=0.14$. The corresponding HSY parameters used in the simulations and MCT calculations are listed in Table \ref{tab:sim_param}. The structure factors required as input
to the MCT method were generated using the analytic MPB-RMSA method. 
It was shown earlier that the MPB-RMSA results for $S(q)$ are in excellent agreement with the MC simulation 
and experimental ones, 
for the concentration series systems referred to in Table \ref{tab:sim_param}. 

\begin{figure}
\vspace{-0.5em}
\includegraphics[width=0.45\textwidth,angle=0]{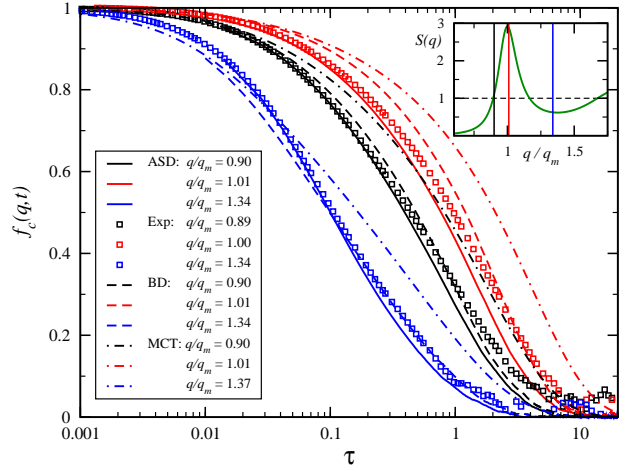}
\vspace{0.0 em}
\caption{\label{fig:fqt_ASD_BD_MCT_EXP}
Normalized intermediate scattering function, $f_c(q,\tau)$, versus reduced time $\tau=t/\tau_a$, for the silica system with $\phi = 0.14$ and three wavenumbers $q$ below (black), at (red) and above (blue) the peak position $q_m$ of $S(q)$. DLS data (open squares) are compared against ASD (solid curves), BD (dashed curves) and MCT (dashed-dotted curves) results. Inset: MC generated $S(q)$, with the vertical lines marking the considered wavenumbers.}
\end{figure}

Fig. \ref{fig:fqt_ASD_BD_MCT_EXP} includes our results for the time dependence of $f_c(q,t)$, 
at $q/q_m = 0.90$ (in black), $q/q_m = 1.01$ near the structure factor peak position (in red), 
and at $q/q_m = 1.34$ near the structure factor minimum (in blue). Solid curves 
are ASD, dashed curves BD, and dashed-dotted curves are MCT predictions. The experimental (DLS) data are presented as open squares.
As discussed 
earlier and seen in the figure, in the short time regime ($\tau \ll 1$) there is quantitative agreement between
the ASD predictions of $f_c(q,\tau)$ and the DLS data. In contrast, the BD and MCT 
results have a different short-time behavior,
owing to their neglect of HIs with the consequence that $H(q)$ is crudely approximated by a constant equal to one. 
At $q/q_m\approx 1.0$ where $H(q)>1$ for the considered low-salinity system,
the short-time decay of the DLS and ASD $f_c(q,\tau)$'s is thus faster than those of the BD and MCT predictions,  Accordingly, the opposite
trend is observed at $q/q_m \approx 1.3$ where $H(q)<1$. At wavenumber $q/q_m\approx 0.9$ where $S(q)\approx 1$ 
(see inset) and $H(q)\approx 1$,
the short-time section of the ASD and BD $f_c(q,\tau)$'s agree with each other as expected, 
and likewise with the DLS data.

At intermediate to long times (i.e., roughly for $\tau \gtrsim 0.1$), the ASD intermediate scattering function 
in Fig. \ref{fig:fqt_ASD_BD_MCT_EXP} follows overall the experimental one, but there is 
no complete quantitative agreement. The ASD $f_c(q,\tau)$ 
decays for $\tau \geq 1$ 
somewhat faster than the experimental one. However, there are unavoidable 
small differences between the ASD and DLS values of the considered three wavenumbers, leading to significant
differences in $f_c(q,t)$ at longer times. Moreover, there are statistical fluctuations in the experimental data, and to a lesser extent
also in the simulation data, blurring the precise wavenumber locations and associated values of the intermediate scattering function.

The MCT and BD results for $f_c(q,\tau)$ can be directly compared, since HIs are disregarded in both cases. This  allows for assessing the accuracy of the MCT method. 
In view of the logarithmic time scale, it is noticed that  
the MCT $f_c(q,t)$ decays substantially slower, at intermediate to long times, than the BD simulation result, implying that the collective width function, $w_c(q,\tau)$, is underestimated using MCT.  
The influence of many-particles HIs follows from the comparison of 
the ASD and BD simulation data, showing that HIs enhances the decay of $f_c(q,t)$ 
at intermediate to long times. 

\begin{figure}
\includegraphics[width=0.45\textwidth,angle=0]{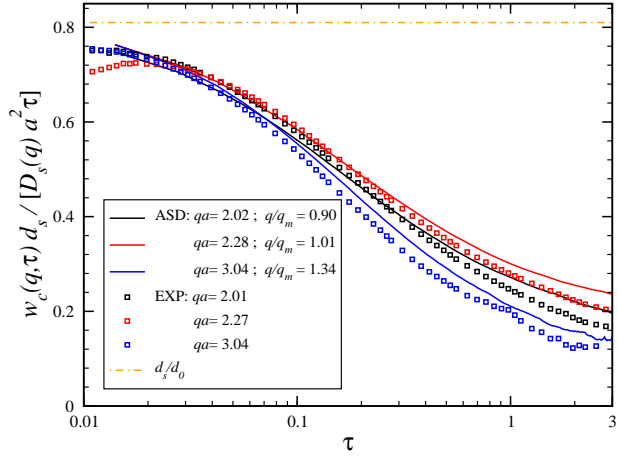}
\vspace{0.0em}
\caption{\label{fig:ASD_EXP_wcqtDsq_t}
Time dependence of the collective width function, $w_c(q,t)/a^2$, divided by its short-time form $D_s(q)t$ and multiplied by $d_0/d_s$, for $\phi = 0.14$.
Three reduced wavenumbers $q/q_m$ are considered as indicated. Open squares are DLS data, 
and solid curves are the ASD simulation results. The dashed-dotted horizontal line marks the ASD value, $d_s/d_0=0.81$, of the reduced short-time self-diffusion coefficient.}
\end{figure}

It is instructive to compare $w_c(q,\tau)$ obtained in the DLS experiments data with the ASD simulation prediction.  For the considered wavenumbers, Fig. \ref{fig:ASD_EXP_wcqtDsq_t} depicts the ratio, $w_c(q,\tau)d_s/\left(D_s(q) a^2\tau\right)$, of the non-dimensionalized collective width function, $w_c(q,\tau)/a^2$, divided by $\tau D_s(q)/d_s$. Since $w_c(q,\tau\ll \tau_a)/a^2 \approx \tau D_s(q)/d_0$, the depicted ratio is referred to in the following as the time-divided width function. At $\tau=0$ it is equal to $d_s/d_0$ which is of value $0.81$ 
according to the ASD simulations (see Table \ref{tab:sim_tcoef}). 
The experimental finding $d_s/d_0=0.77$ listed in Table \ref{tab:fitvalues} is slightly smaller than the ASD value.
Considering the statistical fluctuations in the experimental and simulation data, there is decent overall agreement between the time-divided width functions from ASD and DLS. 

Before analyzing the validity of the time - wavenumber factorization aspect of dynamic scaling using our simulation and MCT results, we scrutinize first whether a long-time exponential decay of $f_c(q,\tau)$ 
is detected in a wavenumber region around $q_m$.

\begin{figure}
\includegraphics[width=0.45\columnwidth,angle=0]{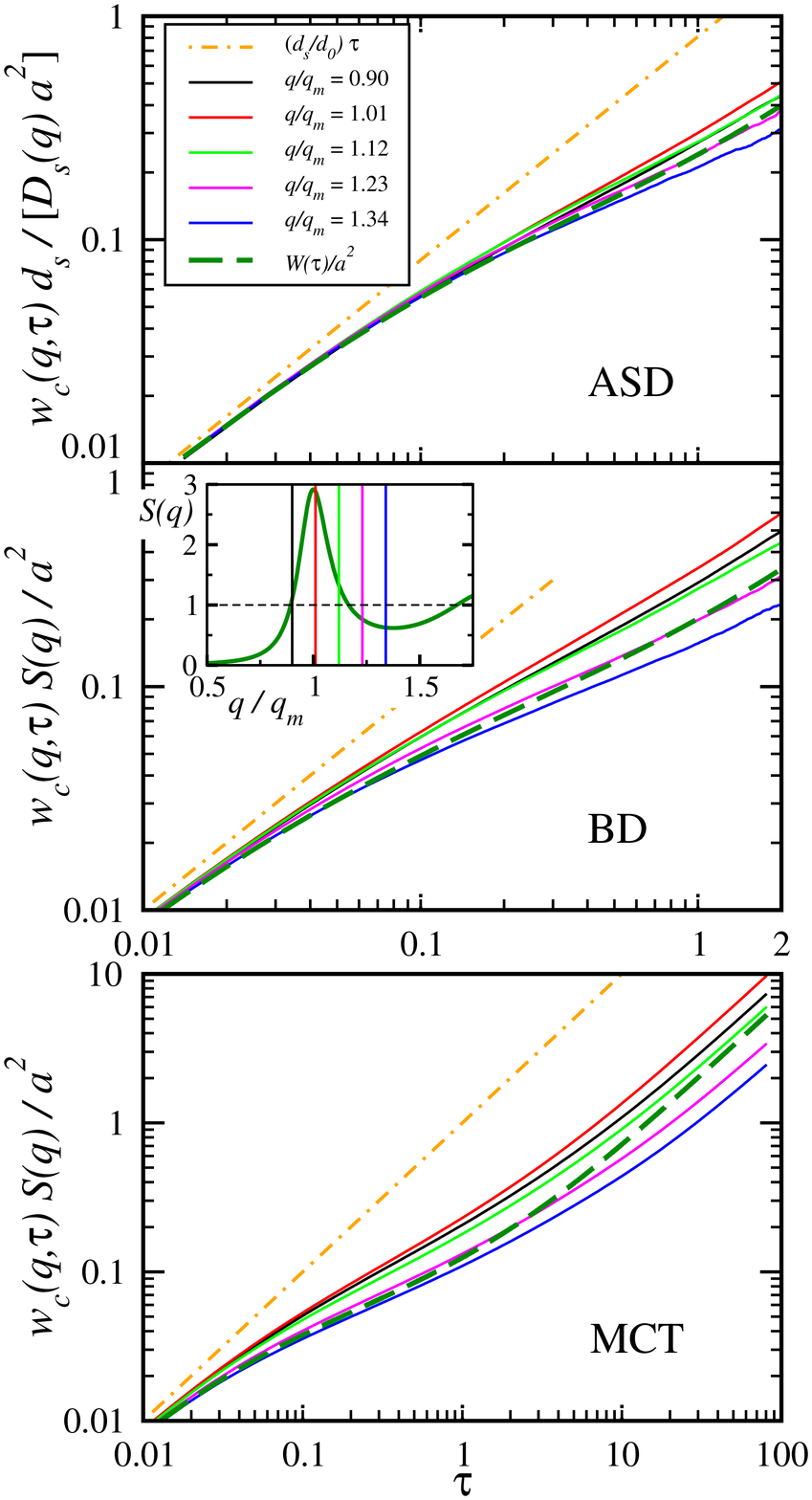}
\vspace{0.0em}
\caption{\label{fig:wc_ASD_BD_MCT}
ASD and BD simulation, and MCT results for the reduced width function 
$(w_c(q,\tau)\;\!d_s)/(D_s(q)\;\!a^2)$, at $\phi=0.14$ and as function of reduced correlation time $\tau$. 
Notice the larger time window of the MCT results in the bottom panel.  
Five wavenumbers $q/q_m$ are considered, distributed in the principal peak region of $S(q)$.
The largest wavenumber is located at the first minimum of $S(q)$ to the right of $q_m$ (see inset showing $S(q)$). 
Also shown is the reduced MSD, $W(\tau)/a^2$, (thick dashed curves) which
increases initially as $(d_s/d_0)\;\!\tau$ (dashed-dotted straight lines). 
Note that $d_s=d_0$ and $D_s(q)=d_0/S(q)$ if HIs are disregarded as
in the BD and MCT calculations. If factorization scaling holds, the non-dimensionalized width function 
is $q$-independent,
and equal to $W(\tau)/a^2$.}
\end{figure}

Fig. \ref{fig:wc_ASD_BD_MCT} displays ASD, BD and MCT results for $w_c(q,\tau)$ (from top to bottom), non-dimensionalized by division through $D_s(q)\;\!a^2/d_s$, revealing its broadening with increasing time. 
The reduced width function increases for short times as $(d_s/d_0)\tau$ (dashed-dotted line) such as the reduced MSD $W(\tau)/a^2$, shown in the figure as the thick dashed green curve. 
Several wavenumbers are considered of values given in the legend box. 
The smallest wavenumber is located to the left of $q_m$ where $S(q)=1$, and the largest one is located at the position of the first minimum of $S(q)$ to the right of $S(q_m)$. 
If HIs are disregarded as in the BD and MCT calculations, the divisor of $w_c(q,t)$ 
is equal to $a^2/S(q)$. For valid factorization scaling, the reduced width function 
should be $q$-independent, and it should agree with $W(\tau)/a^2$. In the
intermediate time range $\tau \sim 0.1-2$ that is well resolved in ASD and BD simulations, 
such a $q$-independent collapse of the width functions is not observed, and likewise not for the MCT data even if only $q_m$ and its closest wavenumbers are examined where $S(q) > 1$. 
Without showing the corresponding curves 
we mention that a narrow
bundle of $w_c(q,\tau)$ curves is found only for wavenumbers selected very close to $q_m$. 
This bundle, however, is located well above $W(\tau)/a^2$.

As a particular feature of dynamic scaling, we explore next whether there is long-time exponential decay of $f_c(q,\tau)$. Provided such a diffusive single-exponential long-time mode is established in the accessed time interval for certain wavenumbers, the collective width function on the double-logarithmic scale in 
Fig. \ref{fig:wc_ASD_BD_MCT} should have two linearly increasing short-time and long-time sections, respectively, 
of slopes equal to one. The short-time section  
is described by $\log(d_s/d_0)+ \log(\tau)$. We discuss here the double-logarithmic plot of the collective width function since such a plot was used and advocated by Martinez {\it et al.} \cite {Martinez2011} to scrutinize experimentally the existence of a long-time exponential decay of $f_c(q,\tau)$ for colloidal hard-sphere suspensions. As described in \cite{Holmqvist2010}, from the DLS data for charge-stabilized silica suspensions the onset of a long-time exponential decay of $f_c(q,t)$ is indeed discernible for strongly interacting spheres (e.g., for $\phi=0.14$) where $d_l$ is significantly smaller than $d_0$, in a wavenumber interval centered at $q_m$. 
From the ASD and BD curves of $w_c(q,t)$ in the present double-logarithmic plot, the onset of a linearly increasing long-time section is discerned only for $q\approx q_m$. 
However, this is no convincing evidence for a long-time diffusive mode 
of $f_c(q_m,\tau)$, since sufficiently long times are not reached in our
simulations without entering  the statistical-numerical noise floor. This limitation is shared by the DLS experiments.

As regards the MCT results for $w_c(q,\tau)$ in the bottom panel of Fig. \ref{fig:wc_ASD_BD_MCT}, 
substantially larger values of $\tau$ are reached extending up to $\tau=100$. In MCT,
an exponential long-time decay of $f_c(q,t)$ is observed for all considered wavenumbers. 
As an independent numerical check, we have obtained this long-time mode in MCT using the singularity analysis of the Laplace transform,  
$\widetilde{f}_c(q,z)$, of $f_c(q,\tau)$ \cite{Banchio:2000} discussed in Subsec. \ref{subsec:Collective dynamics}.  
While MCT predicts a long-time
diffusive mode, no $q$-independent collapse of the MCT curves for $w_c(q,\tau)\;\!S(q)/a^2$ 
on a single one is found for different $q$ values. 
Additionally to predicting clear violation of $q$-$\tau$ factorization of $w_c(q,t)$  
even at $q=q_m$ where $f_c(q,\tau)$ decays most slowly, the MCT     
$w_c(q_m,\tau)\;\!S(q_m)/a^2$ is located well above the MCT curve for $W(\tau)/a^2$.

The non-validity of time-wavenumber factorization scaling of $w_c(q,t)$ predicted by MCT  
does not rule out that this scaling applies approximately, 
for the comparison with the BD data has shown that MCT applies only qualitatively to low-salinity HSY systems. 
An important observation in Fig. \ref{fig:wc_ASD_BD_MCT} is that the spreading of width functions for different $q$ values is less pronounced for the ASD data shown in the top panel. 
Thus, HIs have the effect of re-establishing  factorization
scaling as an approximate feature of $f_c(q,t)$, in the time range accessed in simulations and experiment. 
This explains why the experimental $f_c(q,\tau)$ in \cite {Holmqvist2012} are to a decent degree consistent with factorization scaling. 

\begin{figure}
\includegraphics[width=0.45\textwidth,angle=0]{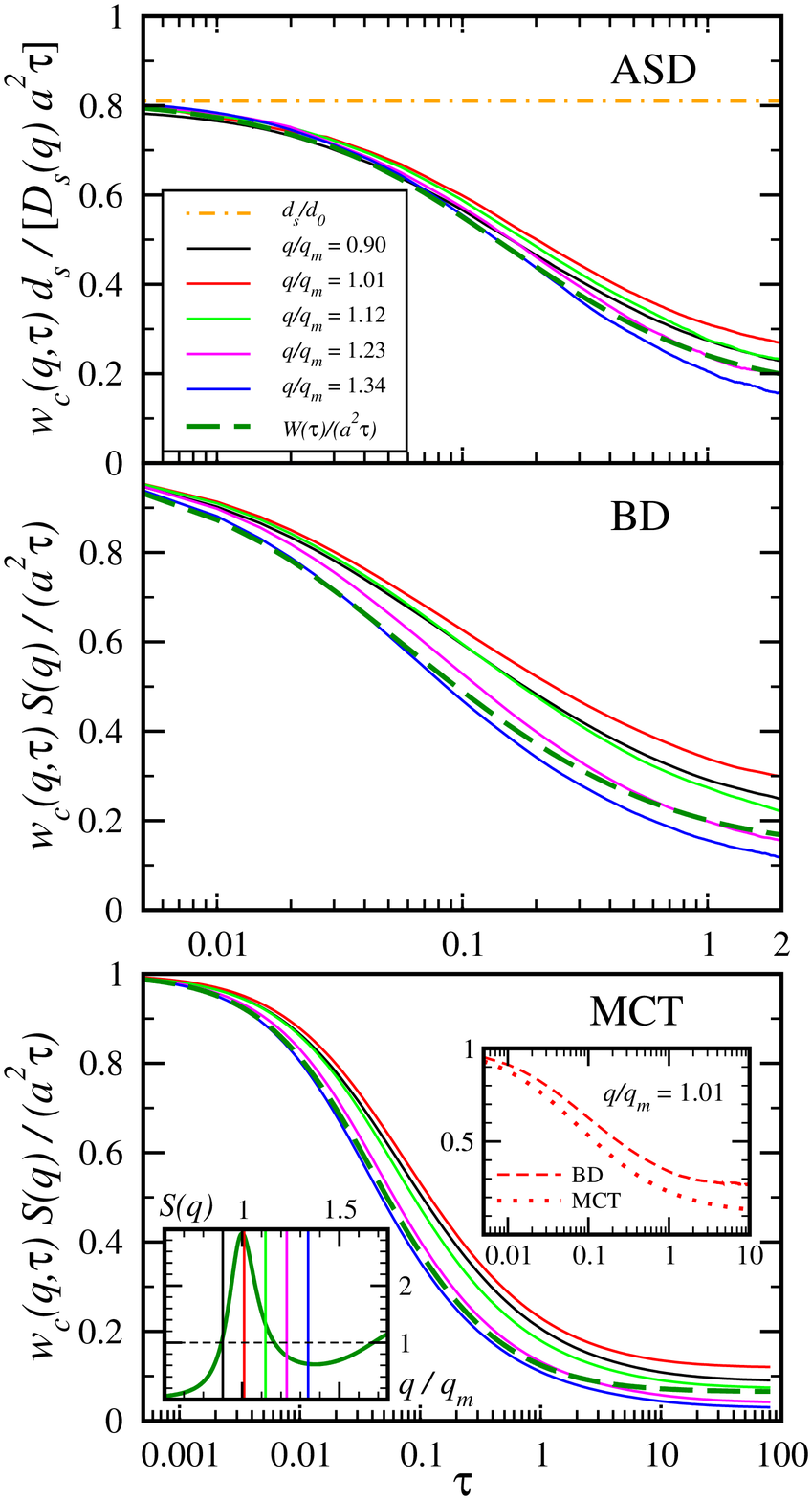}
\vspace{0em}
\caption{\label{fig:wc_t_ASD_BD_MCT}
Time-divided reduced collective width-function, $w_c(q,t)/(D_s(q)t)\times(d_s/d_0)=w_c(q,\tau)d_s/(D_s(q)a^2\tau)$,
for $\phi = 0.14$ and reduced wavenumbers $q/q_m$ given in the top panel, and indicated by vertical lines intersecting $S(q)$ in the inset of the lower panel.
Depicted are ASD, BD and MCT results, with $D_s(q)=d_0/S(q)$ and $d_s=d_0$ in the lower two panels where HIs are disregarded. For valid factorization scaling,
the time-divided width functions should coincide with $W(t)/(d_0 t)=W(\tau)/(a^2\tau)$ (thick dashed green curves).
The inset in the bottom panels compares BD (red dashed) and MCT (red dotted) data at $q/q_m=1.01$.}
\end{figure}
Fig. \ref{fig:wc_t_ASD_BD_MCT} shows the time-divided collective width-function, $w_c(q,t)/(D_s(q)t)\times(d_s/d_0)$, 
on a semi-logarithmic scale, for the same wavenumbers as in the foregoing figure. Note first from the comparison of BD and MCT results at $q\approx q_m$ in the bottom panel shows that MCT significantly underestimates 
collective diffusion at longer times, i.e. the width function value at wavenumber $q_m$ corresponding to the extension, $\sim 2\pi/q_m$,
of the nearest neighbor cage is underestimated. 
For valid $q$-$t$ factorization, 
the time-divided width function curves should agree with the time-divided MSD $W(t)/(d_0 t)$ having the long-time asymptotic value $d_l/d_0$. 
The spread in the time-divided width functions depicted in the figure is least pronounced for the ASD results where many-body HIs are included. 
At longer times, the width function at $q = 1.23\;\!q_m$ has the least deviations from the MSD. Fig. \ref{fig:wc_t_ASD_BD_MCT} shows furthermore, 
in view of the MCT results in the bottom panel where a four times broader time window ($\tau < 100$) is used 
that a genuine long-time plateau of $w_c(q,t)/t$ is not reached within the ASD and BD simulation time windows. 
To deduce the long-time diffusion function $D_l(q)$, if existent, it would be thus profitable to have the time derivative of $w_c(q,t)$ at disposal since $dw_c(q,t)/dt$ converges distinctly faster to $D_l(q)$ than  $w_c(q,t)/t$. While the time derivative is easily computed using MCT, the numerical cost for its calculation is particularly high when system-size corrected ASD simulations are performed. With a very large numerical effort we have managed to generate low-noise numerical results for $dw_c(q,t)/dt$ in an extended time interval and all considered wave numbers. 

\begin{figure}
\includegraphics[width=0.45\textwidth,angle=0]{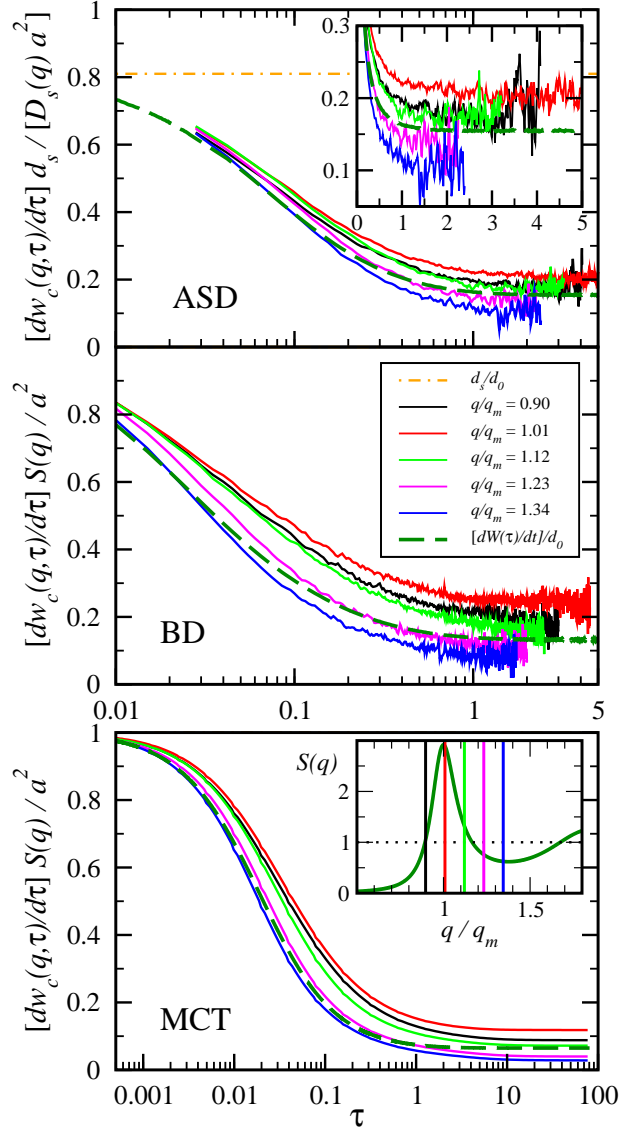}
\vspace{0em}
\caption{\label{fig:dwc_dt_ASD_BD_MCT}
Comparison of ASD (top panel), BD (middle panel) and MCT (bottom panel) results for the
reduced numerical derivative, $dw_c(q,t)dt / \left(D_s(q)t)\times(d_s/d_0)\right)$, 
of the collective width-function, 
for $\phi = 0.14$ and reduced wavenumbers, $q/q_m$, as indicated in the legend and illustrated in the inset of the lower panel.
Dashed thick green curve: Time dependent self-diffusion function, $d(t)/d_0 = (dW(t)/dt)/d_0$, with long-time asymptotic value $d_l/d_0$.
The inset in the top panel depicts ASD data on a linear scale.}
\end{figure}

Simulation and MCT results for the time derivative of $w_c(q,t)$ are displayed in Fig. \ref{fig:dwc_dt_ASD_BD_MCT}. 
For valid factorization scaling, the depicted reduced width function derivative curves should be $q$-independent and coincide with the curve of $d(t)/d_0$, 
with the latter depicted in the figure in green dashed style. 
While the simulation results for $d\;\!w_c(q,t)/d\;\!t$ 
are distinctly noisier at longer times 
than the simulation data for $d(t)$ discussed in the following subsection, there is evidence for the onset of a plateau region for $dw_c/dt$ at the considered wavenumbers.  
We are thus in the position to estimate $D_l(q)/D_s(q)\times (d_s/d_0)$, 
and hence $D_l(q)$, using the depicted simulation data for $dw_c(q,t)/dt$ evaluated for $\tau > 1-2$.
 Results for $D_l(q)$ are discussed at the end of the following subsection on self-diffusion properties. 

\begin{figure}
\includegraphics[width=0.45\textwidth,angle=0]{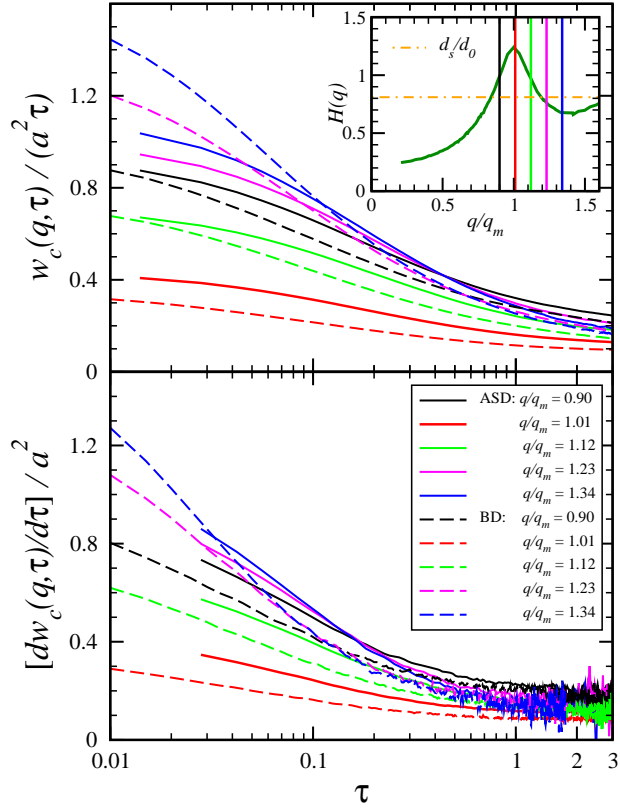}\\
\vspace{0em}
\caption{\label{fig:Figure13}
Time-divided reduced collective width function $w_c(q,\tau)/(a^2\tau)$ (upper panel) and its numerical derivative,
$(dw_c(q,\tau)d\tau)/a^2$ (lower panel), 
for $\phi = 0.14$ and reduced wavenumbers as indicated in the legend and illustrated in the inset, which
depicts the corrected $\delta\gamma$ result for $H(q)$ versus $q/q_m$, reproduced from Fig. \ref{fig:Hq}.}
\end{figure}

We complete our discussion of collective diffusion by analyzing the influence of the HIs.
To this end, Fig. \ref{fig:Figure13} includes a direct comparison of ASD and BD results for
$(w_c(q,t)/(d_0t)$ and $(dw_c(q,t)/dt)/d_0$ in the upper and lower panels, respectively.
Both functions share the same short-time and long-time limits $D_s(q)/d_0$ and $D_l(q)/d_0$.
As seen in the upper panel, for non-small times the ASD generated width function curves with
HIs assume larger values than the corresponding BD curves. It is only for wavenumbers $q$ sufficiently
distant from $q_m$ where $H(q)<1$ and $D_s(q)<d_0/S(q)$, that corresponding ASD and BD curves intersect
at a small time value $\tau \sim 0.01-0.05$. This behavior of the collective width function is
reflected in the lower panel showing $dw_c/dt$. The latter derivative is the $q$-dependent collective
diffusion function analogue of the self-diffusion function $d(t) =\dot{W}(t)$ addressed in the
following subsection. As seen, the lowest collective diffusion function curve is the one for $q=q_m$
where $S(q)$ is largest, and the largest one is for $q=1.34\times q_m$ where $S(q)$ is lowest in the
considered wavenumber interval. Note here again the indication for a plateau region of $dw_c/dt$
at times $\tau \gtrsim 2$.
In closing the discussion of Fig. \ref{fig:Figure13}, we can state for charge-stabilized suspensions
of lower salt content that there is hydrodynamic enhancement of collective diffusion for non-small
correlation times. In the following, we show that HIs likewise enhance self-diffusion. 

%e

\subsection{Self-diffusion at intermediate and long times}\label{sec:sub:Self-diffusion}

\begin{figure}
\includegraphics[width=0.45\textwidth,angle=0]{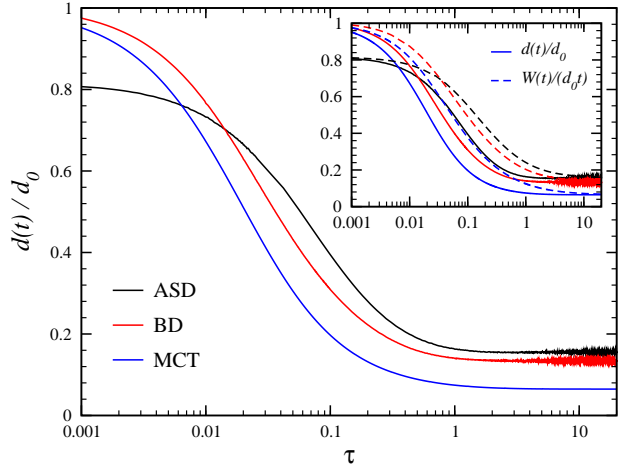}
\caption{\label{fig:dt_ASD_BD_MCT}
ASD, BD and MCT results for the time dependent self-diffusion function, $d(t)=dW(t)/dt$, for $\phi = 0.14$ and depicted in the time range $t \leq 20\;\!\tau_a$. The inset highlights the in comparison with $d(t)/d_0$ much slower decay of $W(t)/(d_0 t)$ (dashed curves) towards the common long-time limit $d_l/d_0$. Hydrodynamic enhancement of self-diffusion at intermediate and long times is observed.
  }
\end{figure}

In Fig. \ref{fig:dt_ASD_BD_MCT}, simulation and MCT results for the normalized time dependent
self-diffusion function, $d(t)=dW(t)/dt$, are plotted as functions of reduced time $\tau$. 
Note from the comparison of ASD and BD results that there is also 
hydrodynamic enhancement of self-diffusion.  
This behavior differs qualitatively from that of suspensions with short-range particle interactions such as colloidal hard-sphere and Hertzian soft sphere suspensions (the latter mimicking non-ionic microgels) 
for which $d_l$ with HIs is smaller than that without HIs \cite{Riest:2015}.   
MCT significantly underestimates $W(t)$ (see inset) and $d(t)$ for $\tau \gtrsim 0.01$.   
In particular the long-time self-diffusion coefficient $d_l$, 
constituting the long-time limit of $d(t)$ and $W(t)/t$, 
is strongly underestimated in MCT (see Table \ref{tab:sim_tcoef}). 

The ASD curves for $d(t)$ and $W(t)$ in Fig. \ref{fig:dt_ASD_BD_MCT} intersect the corresponding BD and MCT 
curves at $\tau \sim 0.01 - 0.02$, since $d_s<d_0$ with HIs. In simulations where the accessible 
time range is limited, 
one preferentially infers $d_l$ from $d(t)$ rather than from $W(t)/t$, since as for $dw_c(q,t)/dt$ the long-time plateau of $d(t)$ is reached at substantially smaller times, i.e.. at $\tau \approx 2 -3$ in Fig. \ref{fig:dt_ASD_BD_MCT}. 
We mention that theory predicts an algebraic long-time tail approach of $W(t)/t$ and $d(t)$ towards $d_l$, 
according to \cite{CichockiHinsen:1992,WagnerHaertl:2001} 
\begin{eqnarray}
 \label{eq:tail}
  W(t)/t &\sim& d_l + \frac{\tau_M\left(d_s - d_l\right)}{t} + {\cal O}(t^{-3/2}) \\
  d(t) &\sim& d_l + {\cal O}(t^{-3/2})\,,
\end{eqnarray}
so that $d(t)$ reaches the common long-time limit $d_l$ much faster than $W(t)/t$, by a factor of ${\cal O}(1/\sqrt{t})$. In principle, an asymptotic tail correction should be made in deducing $d_l$ from simulation results covering only a limited time range. Chichocki and Hinsen in \cite{CichockiHinsen:1992} performed such a long-time correction to infer $d_l$ from their BD simulation data for concentrated hard-sphere suspensions without HIs. For this purpose, they used a so-called two-pole approximation for the relaxation function $\gamma(t)$ in Eq. (\ref{eq:self-diffusion-function}). Here, $\tau_M$ is the mean relaxation time of particle velocity auto-correlations that characterizes likewise the decay of $d(t)$. We have refrained from an asymptotic tail analysis since our simulation results cover a distinctly broader time range than the earlier hard-sphere simulations by Chichocki and Hinsen which extend up to $\tau = 2.5$ only. Furthermore, as discussed we have succeeded in calculating $d(t)$ to high accuracy in this extended time range. 
For the charged-particles suspensions, we have derived an analytic expression 
for $\gamma(t)$ (not shown here), valid in the weak coupling limit. This expressions conforms with Eq. (\ref{eq:tail}), and it identifies the characteristic decay time of $d(t)$ and $\gamma(t)$ in this limit as $\tau_M=1/(2\kappa^2d_0)$ where $\kappa$ is the electrostatic screening parameter.

\begin{figure}
\vspace{3.5em}
\includegraphics[width=0.45\textwidth,angle=0]{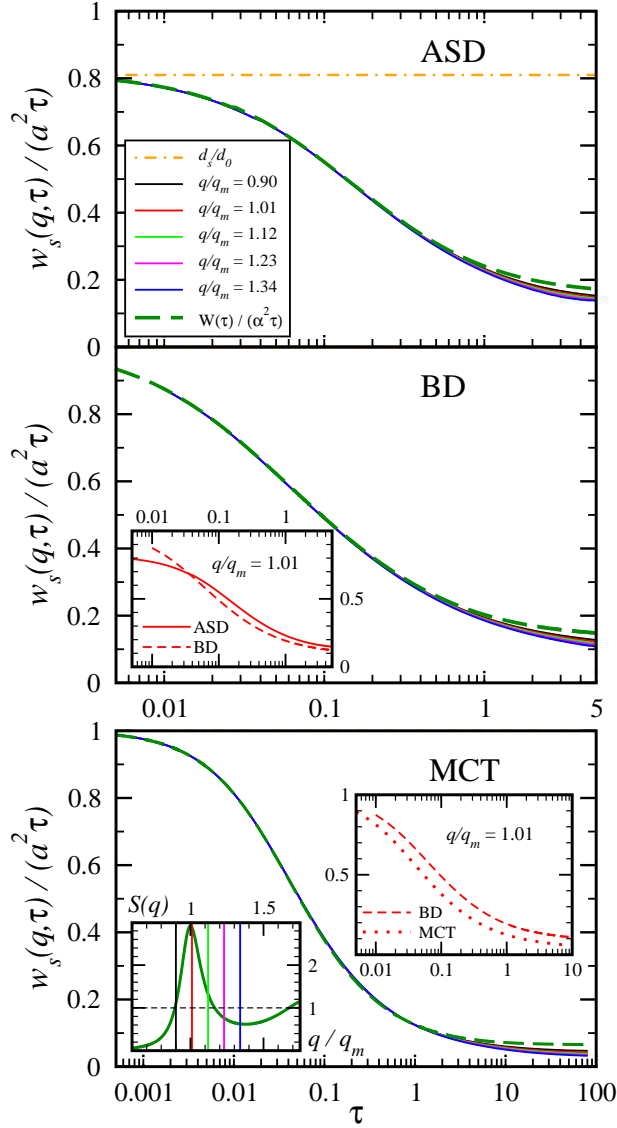}
\vspace{0em}
\caption{\label{fig:ws_over_t_ASD_BD_MCT}
ASD, BD and MCT results for the time-divided self-diffusion width function, $w_s(q,t)/(d_0 t)=w_s(q,\tau)/(a^2 \tau)$, (solid curves)
as functions of reduced time $\tau$, and for $\phi = 0.14$. The considered reduced wavenumbers, $q/q_m$, are indicated by vertical lines
intersecting $S(q)$ in the left inset of the lower panel. The time-divided width function is equal to $d_s/d_0$ at $t=0$, and without non-Gaussian effects it is
identical to $W(t)/d_0 t=W(\tau)/(a^2\tau)$ (thick dashed green curve). 
Red curves in the insets of the middle and bottom panel are ASD (solid), BD (dashed) and MCT (dotted) results, all taken at $q/q_m=1.01$. 
}    
\end{figure}

We proceed with discussing the influence of non-Gaussian effects on the self-diffusion width function $w_s(q,t)$ introduced in Eq. (\ref{eq:fsq}). Without these effects, $w_s(q,t)$ would be $q$-independent and equal to $W(t)$. 
However, $w_s(q,t)= W(t)$ holds strictly for $q \ll q_m$ only.  
Fig. \ref{fig:ws_over_t_ASD_BD_MCT} displays our theoretical results for the time-divided self-diffusion width function, $w_s(q,t)/(d_0 t)$ (solid curves), on a log-linear scale and for different
$q$-values. These results are compared against $W(t)/(d_0 t)$ (thick green dashed curve), 
to globally identify non-Gaussian effects causing the time-divided width function to differ from $W(t)/(d_0 t)$. The initial value of all displayed curves is equal to $d_s/d_0$. 
Quite interestingly, non-Gaussian effects are 
visible in this plot for larger times only. The deviations are most pronounced at the largest considered wavenumber, 
$q = 1.34\times q_m$, at a structure factor minimum for which the  MCT calculated curve of $w_s(q,t)/(d_0 t)$ 
is for $\tau > 10$ well below the MCT $W(t)/(d_0 t)$. In the inset of the middle panel, 
the ASD and BD time-divided self-diffusion width functions for $q/q_m =1.01$ are directly compared against each other. 
Akin to the self-diffusion function in Fig. \ref{fig:dt_ASD_BD_MCT}, the ASD and BD curves intersect at $\tau \approx 0.02$, 
with a subsequent hydrodynamic enhancement of the self-diffusion width function for $\tau \gtrsim 0.02$. 
According to the inset in the bottom panel where BD and MCT results at $q/q_m = 1.01$ are compared, 
the slowing influence of particle caging caused by direct interactions is overestimated using MCT. 

\begin{figure}
\vspace{-0.5em}
\includegraphics[width=0.45\textwidth,angle=0]{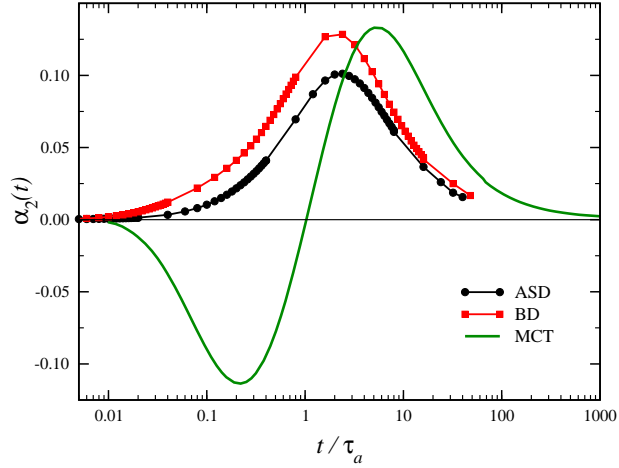}
\vspace{0 em}
\caption{\label{fig:NonGaussian}
Non-Gaussian parameter, $\alpha_2(t)$, for $\phi=0.14$ according to ASD, BD and MCT as indicated.   
For fluid state systems such as the considered one the long-time limiting value is $\alpha_2(t \to \infty)=0$. 
Furthermore, $\alpha_2(0)=0$ without HIs 
while (near-field) HIs induce a small variance in the longitudinal hydrodynamic self-mobility, with $\alpha_2(0)$ being consequently slightly positive valued.
  }
\end{figure}

To quantify non-Gaussian effects in self-diffusion, we determine the leading-order contribution, $\alpha_2(t)$, of the non-Gaussian parameters in three dimensions,
\begin{eqnarray} \label{eq:alpha}
    \alpha_n(t)=c_n\times\frac{\big<\left|{\bf r}_i(t)-{\bf r}_i(0) \right|^{2n}\big>}{\big<\left|{\bf r}_i(t)-{\bf r}_i(0) \right|^{2}\big>^n}\;\!-\;\!1\,,
\end{eqnarray}
where $c_n=3^n/(2n+1)!!$ for $n=\{2,3,\cdots\}$ with $c_2=3/5$. These parameters are the coefficients in the cumulant expansion \cite{WagnerHaertl:2001,vanMegenSnook1981},
\begin{eqnarray}
       f_s(q,t) =\exp\Big\{-q^2\;\!W(t) +\frac{\alpha_2(t)}{2}(q^2\;\!W(t))^2 + \frac{3\alpha_2(t) -\alpha_3(t)}{3!}(q^2\;\!W(t))^3 + \cdots \Big\}\,,
\end{eqnarray}
of the logarithm of $f_s(q,t)$ in powers of $q^2 W(t)$. 
If the particle displacement of a particle $i$ is a Gaussian process, all $\alpha_n(t)$'s are identically zero, and $f_s(q,t)$ is equal to $\exp\{-q^2\;\!W(t)\}$. Before showing the results for $\alpha_2(t)$, we address its general properties. Since the fourth-order displacement moment, $\big<\left|{\bf r}_i(t)-{\bf r}_i(0) \right|^{4}\big>$, in the nominator of Eq. (\ref{eq:alpha}) for  $n=2$ is larger or equal to the second moment in the denominator, 
it holds that $\alpha_2(t) \geq -2/5$, i.e. only negative values above this lower bound are allowed. From the definition of $\alpha_2(t)$ as the coefficient of the $q^4$ term in the cumulant expansion, and from the self-diffusion memory Eq. (\ref{eq:memory_eq_self}) for $f_s(q,t)$, it follows for a fluid system where $M_s^{irr}(q,t\to \infty)=0$ that $\alpha_2(t\to \infty) =0$. In the short-time regime, the exact expression applies \cite{Tough:1986},
\begin{eqnarray}\label{eq:Tough}
  \alpha_2(t) = \left[\frac{\big< {\left(\hat{{\bf q}}\cdot \boldsymbol{\mu}_{11}\cdot \hat{{\bf q}}\right)}^2\big>}
  {{\big< {\hat{{\bf q}}\cdot \boldsymbol{\mu}_{11}\cdot \hat{{\bf q}}} 
  \big>}^2} - 1\right] + {\cal O}(t)\,,
\end{eqnarray}
where $\boldsymbol{\mu}_{11}({\bf r}^N)$ is the self-mobility tensor and $\big< \hat{{\bf q}}\cdot \boldsymbol{\mu}_{11}\cdot \hat{{\bf q}}\big>=d_s/(k_BT)$. According to this expression, 
there is a non-Gaussian contribution $\alpha_2(0)>0$ already at $t=0$, 
provided the near-field part of the HIs is significant. The variance in the longitudinal hydrodynamic self-mobility, 
$\hat{{\bf q}}\cdot \boldsymbol{\mu}_{11}\cdot \hat{{\bf q}}$, and hence $\alpha_2(0)$, are zero for systems where only the far-field (i.e. Rotne-Prager) part of the HIs influences the dynamics  so that $d_s=d_0$, or when HIs are fully neglected such as in our free-draining BD simulations. Non-Gaussian corrections due to direct (non hard-sphere) particle interactions in $\alpha_2(t)$ appear first to order $t^2$  \cite{Tough:1986}. 

Simulation results for $\alpha_2(t)$ obtained from direct calculation of the quartic displacement moment are shown in Fig. \ref{fig:NonGaussian}, in comparison with the MCT prediction. In MCT, the quartic moment can be calculated as the numerical solution of an integro-differential equation derived from singling out the $q^4$ expansion term in the MCT equation for $f_s(q,t)$ \cite{FuchsGaussian:1998,FlennerSzamel:2005}. Since the structure factor input to the MCT is efficiently calculated using the MPB-RMSA method, we follow an alternative route by determining $\alpha_2(t)$ from a linear (in terms of $q^2$) small-wavenumber fit of $(f_s(t,t)-1)/q^2$ according to
\begin{eqnarray}
  \frac{f_s(q,t)-1}{q^2} \approx  - W(t) +\frac{(1 +\alpha_2(t))W^2(t)}{2}\;\! q^2 W(t)\,. 
\end{eqnarray}
\begin{table}
%{
\footnotesize{ % Just for OneColumn
\caption{ASD, BD and MCT values for the long-time diffusion function $D_l(q_m)$, for $\phi=0.14$. Regarding ASD and BD, only an estimate is deduced operationally from the numerical derivative $dw_c(q_m,t)/dt$, 
considered at the largest accessed time values where statistical fluctuations are still small. 
Values for $D_s(q_m)$, $d_s$ and $d_l$ are taken from Tables \ref{tab:sim_tcoef} and \ref{tab:fitvalues}. In BD and MCT is $d_s=d_0$ and $D_s(q)=d_0/S(q)$. For valid dynamic scaling is $D_l(q_m)/D_s(q_m)\times d_s/d_l=1$.\\
} \label{tab:Dlong}}
\vspace{.1em}
\centering
{\footnotesize{
\begin{tabular}{@{\extracolsep{\fill}}llll}
\hline
$\boldsymbol{}$\quad~&
$\mathrm{D_l(q_m)/d_0}$\quad~&
$\mathrm{D_l(q_m)/D_s(q_m)}$\quad~&
$\mathrm{D_l(q_m)/D_s(q_m) \times d_s/d_l}$\\
\hline\hline
\text{ASD\qquad}  &  0.10  & 0.21  & 1.1  \\
\text{BD}   &  0.09  & 0.25  & 2.0  \\
\text{MCT}  & 0.04   & 0.12  & 1.8  \\
\hline
\end{tabular}
}}
\end{table}
The ASD and BD simulation results for $\alpha_2(\tau)$ in Fig. \ref{fig:NonGaussian} are non-negative and monotonically increasing for smaller $\tau$, 
passing through a maximum at $\tau \approx 2$ where $d(\tau)$ in Fig. \ref{fig:dt_ASD_BD_MCT} has nearly reached its long time value $d_l$. 
For further increasing $\tau$, the non-Gaussian parameter decreases rather slowly towards its long-time asymptotic value zero. 
Comparison of ASD and BD results for $\alpha_2(t)$ reveals the effect of HIs in mitigating non-Gaussian contributions arising from the screened Coulomb interactions. 
In ASD simulations where HIs are included the initial value $\alpha_2(0)$ is slightly positive as described by Eq. (\ref{eq:Tough}). Our ASD result is indeed indicative 
of a small positive value not resolved on the scale of the figure. 
The time dependence of $\alpha_2(t)$ according to MCT differs qualitatively from those of the simulation results, 
since for $\tau \lesssim 1$ negative values are predicted. While the positive-valued maximum of the MCT $\alpha_2(t)$ is of similar height than that of the BD $\alpha_2(t)$, 
its location is shifted to a larger value $\tau \sim 3$, followed by the expected monotonic decline towards $\alpha_2(\infty)=0$. 
The MCT initial value $\alpha_2(t=0)$ is equal to zero within numerical accuracy  
as required without HIs.  
While the exact constraint $\alpha_2(t) > -0.4$ is fulfilled here by the MCT non-Gaussian parameter, 
this is in general not guaranteed using this approximative method \cite{FuchsGaussian:1998}. 
The deficiency of the MCT can be attributed to the fact \cite{FuchsGaussian:1998,Banchio:2000} 
that the self- and collective memory function approximations by this method are more severe 
for short and intermediate times where negative values 
of $\alpha_2(t)$ are observed here conflicting with the BD prediction. Non-Gaussian contributions to $f_s(q,t)$ are rather small for the considered HSY system. However, when a glass transition point is approached, 
the maximum of $\alpha_2(t)$ can attain 
significantly larger values than observed here 
\cite{FuchsGaussian:1998,FlennerSzamel:2005}.            
 
\begin{figure}
\includegraphics[width=0.45\columnwidth,angle=0]{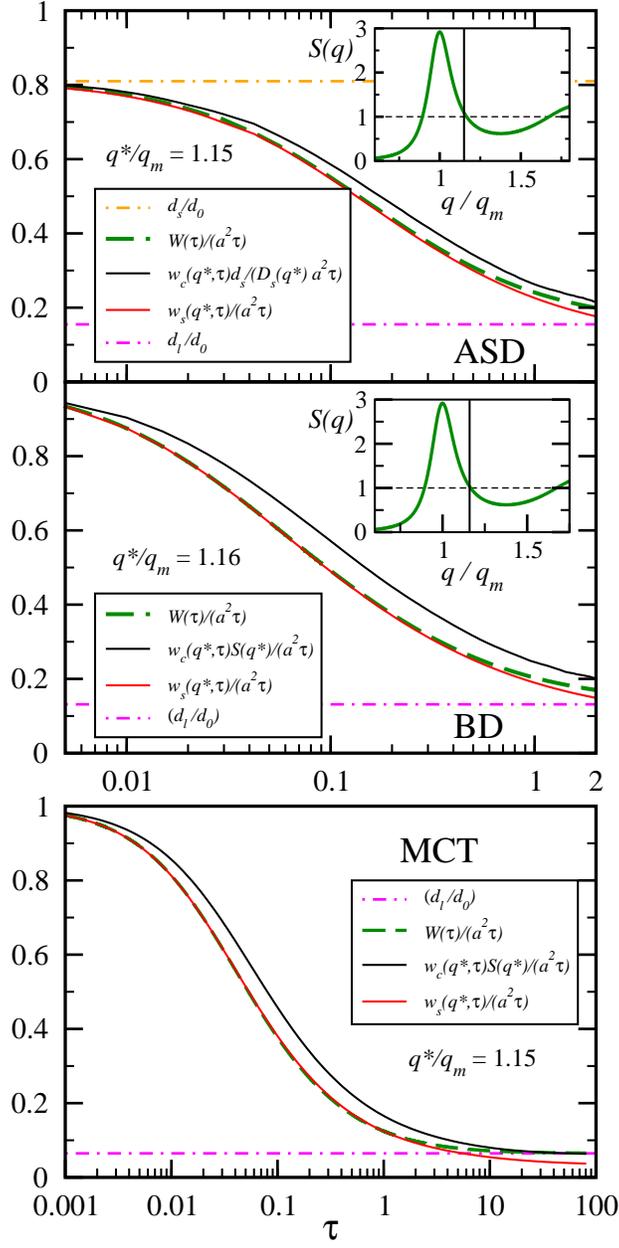}
\vspace{0em}
\caption{\label{fig:self_from_coll_dt}
ASD, BD and MCT results for the time-divided MSD $W(t)/(d_0 t)$ (green dashed curve), and the collective and self-diffusion time-divided reduced width functions, $w_c(q^\ast,t)/(D_s(q^\ast)t)\times(d_s/d_0)$ (black curve) and $w_s(q^\ast,t)/(D_s(q^\ast)t)\times(d_s/d_0)$ (red curve), at the special wavenumber $q^\ast=(1.15-1.16)\times q_m$ where $S(q^\ast)\approx 1$ (see inset) and $D_s(q^\ast) \approx d_s$. The system with $\phi = 0.14$ is considered.
All curves start at $d_s/d_0$ for $t=0$, of value marked in top ASD panel by the upper dashed-dotted yellow line.
The lower dashed-dotted violet line gives $d_l/d_0$. In the ASD and BD simulations is $w_s(q^\ast,\tau) \approx W(\tau)$ valid up to intermediately large times. The collective width function, $w_c(q^\ast,\tau)$, tends to overestimate the MSD.
}
\end{figure}

As noted in Subsec. \ref{sec:sub:Results_shorttime} in the context of short-time collective diffusion, for many experimental suspensions it is difficult to deduce self-diffusion properties directly from a standard DLS measurements of $f_c(q,t)$. In extending Pusey's approach \cite{Pusey1978, Segre1995, Abade2010} to intermediate and long times, one may try to infer $W(t)$ and $d_l$ approximately from a DLS measurement of $f_c(q,t)$, and hence of $w_c(q,t)$, performed at a wavenumber $q^\ast>q_m$ where $S(q^\ast)\approx 1$ and $D_s(q^\ast)\approx d_s$. As a theoretical check   of the accuracy of this approach, Fig. \ref{fig:self_from_coll_dt} compares 
the time-divided reduced collective width function, $w_c(q^\ast,t)/(D_s(q^\ast)t)\times(d_s/d_0)$, with the time-divided MSD, $W(t)/(d_0 t)$, and likewise with the time-divided self-width function $w_s(q^\ast,t)/(d_0 t)$. The wavenumber $q^\ast$ is usually located inside the experimental wavenumber range. If Pusey's approach remains useful at longer times, $W(t)\approx w_c(q^\ast,t)$ should be a reasonable approximation of the MSD. According to the figure, the time-divided collective width function is visibly larger than $W(t)/(d_0 t)$, with the long-time limit $d_l/d_0$ of the latter marked in the figure by the lower horizontal dashed-dotted line. Quite interestingly, the difference between the collective width function at $q^\ast$ and the MSD is smaller with HIs than without it. Thus, while there is no quantitative agreement, DLS measurements of $f_c(q,t)$ at $q^\ast$ are seen here to be useful in obtaining 
reasonable estimates of $W(t)$, and an upper bound of $d_l$. The good agreement between 
$w_s(q^\ast,t)/(D_s(q^\ast)t)\times(d_s/d_0)$ and $W(t)/(d_0 t)$ 
in the ASD and BD panels, respectively, points to the smallness 
of non-Gaussian effects at the considered wavenumber $q^\ast$ 
and the intermediate time range. 
 
Having addressed the determination of self-diffusion properties including $d_l$, we are now in the position 
to scrutinize the dynamic scaling prediction regarding the long-time diffusion function $D_l(q)$. Recall here  the evidence for long-time exponential decay of $f_c(q,t)$ gained indirectly from the observed decay of $dw_cq,t)/dt$ towards a long-time plateau value. There is additional theoretical evidence 
coming from MCT results for charged and neutral colloidal particles, and from the CEA method for hard spheres  which unanimously predict a long-time mode in particular at $q=q_m$, provided the particles are sufficiently strongly correlated such as in the present case where $S(q_m)=2.9$ at $\phi=0.14$. In Table \ref{tab:Dlong}, values for $D_l(q_m)$ are given, obtained in an operational way from considering the numerical derivative $dw_c(q_m,t)/dt$ at the largest accessed correlation times. For valid dynamic scaling,  
this derivative is equal to $d(t)\times d_s/D_s(q_m)$. 
Regarding ASD and BD, 
the listed values for $D_l(q_m)$ are only estimates,  
since in the plateau region $d\;\!w_c(q_m,t)/d\;\!t$ fluctuates quite strongly. 
While the dynamic scaling prediction 
$D_l(q_m)/D_s(q_m)\times(d_s/d_l)=1$ 
is clearly violated by the BD and MCT data 
where HIs are neglected, 
a value close to one 
is obtained from the ASD simulations. 
However, the good agreement between ASD result 
and dynamic scaling prediction should be taken 
with a pinch of salt, 
considering the 
significant statistical fluctuations 
of the ASD data at longer times.       

\section{Conclusions}\label{sec:Conclusions}

We have presented the first comprehensive study of collective and self-diffusion properties of charge-stabilized suspensions. For this purpose, 
elaborate Brownian particles simulations without and with full many-particles HIs included were combined with MCT scheme calculations, and DLS/SLS/SAXS experiments on low-salinity suspensions of charged silica spheres in an organic solvent mixture.

In simulations and theoretical calculations, 
we used the HSY model of charge-stabilized colloidal spheres. The effective HSY potential parameters $Z$ and $\kappa$ were  determined from fits of the experimental SLS $S(q)$, using MC simulations, numerical solutions of the RY scheme, and the MPB-RMSA method. The excellent structure factor fits obtained by these methods for identical HSY parameters demonstrates, first, that the analytic MPB-RMSA method provides reliable structure factors required as input by the MCT method in a  broad $q$-range. Secondly, the pair structure of the considered suspensions of nearly monodisperse silica spheres is well described by the HSY model. That the HSY model is an appropriate description is additionally shown by the good agreement between the short-time DLS data for $D_s(q)$, and the ASD predictions for this function using the same HSY parameters as in the $S(q)$ calculations. A result of practical relevance is that 
the elaborate ASD results for $D_s(q)$ and $H(q)$ are well reproduced by the self-part corrected $\delta\gamma$ scheme with MPB-RMSA structure factor input. It was shown recently that this hybrid scheme for short-time diffusion properties is well suited also for dispersion of particles showing comparably strong short-range 
attraction and long-range repulsion, such as for low-salinity globular protein solutions \cite{RiestSALR:2015,Shiba:workinprogress}.      
 
Our ASD and BD simulation results include $f_c(q,t)$ and $f_s(q,t)$ and their associated width functions, for a set of wavenumbers encompassing the principal peak region of $S(q)$. Additionally, the self-diffusion properties $W(t)$, $d(t)$, $d_s$ and $d_l$ were studied. Regarding the simulations, a generalized system size correction  procedure was introduced applying for all correlation times. Since the statistical errors in the simulation results for the self-diffusion properties are smaller than for the collective ones, 
we succeeded to obtain reliable simulation results for the time derivative, $d(t)$, of the MSD decaying substantially faster than $W(t)/t$ towards the common long-time plateau value $d_l$. This allowed us to infer the long-time self-diffusion coefficient from the simulations. 

We further assessed quantitatively how a decent estimate of the MSD $W(t)$ at intermediate times is obtained from $f_c(q^\ast,t)$ 
considered at a wavenumber $q^\ast >q_m$ where $S(q^\ast)=1$, and we quantified non-Gaussian effects on the self-intermediate scattering function. In particular, we showed that HIs 
lessen the leading-order non-Gaussian parameter $\alpha_2(t)$. 
The comparison of ASD and BD simulation results revealed that collective diffusion, as quantified by $dw_c(q,t)/dt$, is at all times enhanced by HIs, for wavenumbers at and near to $q_m$ where $H(q)>1$. Self-diffusion, as quantified by $w_s(q,t)$ and $d(t)$, 
is likewise hydrodynamically enhanced at intermediate and long times, but not at short times since  $d_s < d_0$. 
The observed enhancement of self-diffusion and collective diffusion is due to the long-range part of the HIs which is strongly influential 
for the considered low-salinity systems where near-contact particle pairs are very unlikely.

The ASD and BD simulations results, and the 
MCT results for $f_c(q,t)$ were compared against our DLS data on a charged silica particles suspension for  
$\phi=0.14$. While at short times the DLS data are in excellent agreement with the ASD 
predictions of $D_s(q)$, except for low-$q$ statistical fluctuations,  
the decay of the DLS $f_c(q,t)$ is mildly overestimated using ASD 
for larger times $\tau \gtrsim 0.5$, i.e. the associated collective width function, $w_c(q,t)$, is mildly overestimated. 
There is good overall agreement between DLS and ASD data, also regarding the time-divided collective width function, $w_c(q,t)/(D_s(q,t)t)$, which is only mildly overestimated by the ASD simulations at longer times. 

We have scrutinized dynamic scaling, a far reaching proposition whose (approximate) validity would allow to infer self-diffusion from collective diffusion properties (in a $q$ interval around $q_m$), and long-time diffusion coefficients and functions from the corresponding short-time ones. 
In the time range where $f_c(q,t)$ is accessed in the ASD and BD simulations, a spreading of the time-divided width function curves, $w_c(q,t)/D_s(q)t$, is observed for the different wavenumbers.  
These curves do not collapse on a $q$-independent master curve equal to that of $W(t)/(d_s t)$, as posited by the time-wavenumber factorization facet of dynamic scaling. Decently good agreement with $W(t)/(d_s t)$ is found for $q/q_m \approx 1,23$ only, while deviations from the time-divided MSD are largest at $q_m$. In view of the ASD simulation results, we conclude that factorization scaling is a better approximation when HIs are accounted for, causing a significantly narrower bundle of collective width curves. 
This explains why for the silica particles systems explored in \cite{Holmqvist2010} using DLS, factorization scaling was noticed to be approximately valid in a wavenumber interval comparable with that considered in this work. 

The long-time exponential decay of $f_c(q,t)$ could not be inferred directly from the ASD and BD simulation results for the collective width function, 
owing to the slow decay of $w_c(q,t)/t$ in the time 
range where the statistical fluctuations in the simulation data stay small. 
However, there is an indication of long-time plateau values of the width function derivative (collective diffusion function) $d w_c(q,t)/d t$
which decays distinctly faster than $w_c(q,t)$.  
This has allowed us to infer simulation estimates of the long-time cage diffusion coefficient $D_l(q_m)$. 
Combined with the simulation results for $d_l$ identified with the long-time plateau value of $d(t)$, 
we scrutinized the long-time factorization prediction $D_l(q_m)/D_s(q_m)=d_l/d_s$ which according to the ASD results is satisfied up to ten percent deviation. 
This supports additionally our conclusion for strongly correlated charged-particles suspensions that 
thanks to HIs, dynamic scaling is valid as an approximate feature. Without depicting explicit results we note that according to simulations and theory alike, $D_l(q)/D_s(q)$ varies non-monotonically as a function of $q$, taking its maximal value larger than $d_l/d_s$ at $q=q_m$, and attaining values smaller than $d_l/d_s$ for $q \gtrsim 1.2\times q_m$.         

MCT results for collective and self-dynamic properties were presented in a broad time range $\tau <100$, 
extending well into the long-time regime of the dynamic scattering functions. 
In comparison with ASD and BD simulation results, MCT predicts too slowly decaying 
scattering functions $f_c(q,t)$ and $f_s(q,t)$ for non-short times, underestimating consequently the associated width functions and $W(t)$. 
Since the overestimation of dynamical particle caging by MCT persist also in comparison with the BD data 
where HIs are likewise disregarded, this shows the limited accuracy of MCT predictions 
for the dynamics of low-salinity HSY systems. 
This is most evident regarding $\alpha_2(t)$ for which negative values are predicted for $\tau \lesssim 1$, 
in qualitative disagreement with our simulation results for the non-Gaussian parameter that are non-negative at all times. 
While in MCT a long-time exponential decay of $f_c(q,t)$ is found for all considered wavenumbers, 
different from ASD a very significant violation of $q$ - $t$ factorization of $w_c(q,t)$ is predicted. 

In future work, we will undertake a comprehensive simulation study of colloidal hard-sphere suspensions, constituting the opposite high-screening limit of the HSY model discussed here for weak screening. 
For colloidal hard-sphere suspensions, high-resolution experimental data for the intermediate and self-intermediate scattering functions, and $W(t)$, have been obtained in particular by van Megen and collaborators (see, e.g.,\cite{Martinez2011,vanMegenSchoepe:2017}). To calculate diffusion properties, 
the CEA method can be then used  
in addition to MCT.   
Both methods can be amended for hard spheres by a short-time rescaling method  \cite{Banchio1999,Banchio:2000,Riest:2015} allowing for an approximate inclusion of HIs. 
Work in this direction is in progress. \\

\section*{acknowledgement}
G. N. thanks W. van Megen (RMIT, Melbourne) for helpful discussions.
M. H. acknowledges financial support from CONACyT (Grant No. 237425/2014).

\section*{Abbreviations}
\renewcommand{\arraystretch}{1.25}
\begin{longtable}{p{.25\columnwidth}p{.75\columnwidth}}
ASD\dotfill&Accelerated Stokesian Dynamics simulations (Brownian particles)\\
BD\dotfill&Brownian Dynamics simulations (without HIs: free draining)\\
CEA\dotfill&Contact Enskog type approximation\\
DLS\dotfill& Dynamic light scattering\\
DLVO\dotfill& Derjaguin-Landau-Verwey-Overbeek (pair potential)\\
HIs\dotfill& Hydrodynamic interactions\\
HS\dotfill& Hard spheres\\
HSY\dotfill& Hard-sphere plus Yukawa (pair potential)\\
MC\dotfill& Monte Carlo simulation\\
MCT\dotfill& Mode-coupling theory (without HIs)\\
MPB-RMSA\dotfill& Modified penetrating-background corrected rescaled mean spherical approximation\\
MSD\dotfill& Mean-squared displacement\\
PA-scheme\dotfill& Approximation of pairwise additive hydrodynamic interactions\\
PNIPAM\dotfill& Poly (N-isopropylacrylamide)\\
RY\dotfill& Rogers-Young (integral equation scheme)\\
SAXS\dotfill& Small-angle x-ray scattering\\
SLS\dotfill& Static light scattering\\
SDF\dotfill& Spectral distribution function of relaxation rates\\
TPM\dotfill& Trimethoxysilylpropyl methacrylate\\
WCL\dotfill& Weak coupling limit of particles interacting by Fourier-integrable $u(r)$\\
XPCS\dotfill& X-ray photon correlation spectroscopy\\
\end{longtable}
~\newline


\begin{thebibliography}{99}

\bibitem{RusselBook:1989}
W.B. Russel, D.A. Saville, and W.R. Schowalter.
\newblock {\em Colloidal Dispersions}.
\newblock Cambridge University Press, New York, 1989.

\bibitem{Nagele1996}
G.~N\"{a}gele.
\newblock {\em Phys. Rep.}, 272:216, 1996.

\bibitem{Likos_MicrogelReview:2011}
C.~N. Likos.
\newblock Structure and thermodynamics of ionic microgels.
\newblock In {\em Microgel Suspensions: Fundamentals and Applications},
  chapter~7. Wiley-VCH Verlag GmbH \& Co. KGaA, 2011.

\bibitem{Holmqvist2012}
P.~Holmqvist, P.~S. Mohanty, G.~N\"agele, P.~Schurtenberger, and M.~Heinen.
\newblock {\em Phys. Rev. Lett.}, 109:048302, 2012.

\bibitem{ivlev2012complex}
Alexei Ivlev, Hartmut L{\"o}wen, Gregor Morfill, and C~Patrick Royall.
\newblock {\em Complex Plasmas and Colloidal Dispersions : Particle-Resolved
  Studies of Classical Liquids and Solids}.
\newblock Series in Soft Condensed Matter - Vol. 5. World Scientific, 2012.

\bibitem{NaegeleEPJST:2013}
G.~N{\"a}gele.
\newblock {\em Eur. Phys. J. Special Topics}, 222:2855, 2013.

\bibitem{SegrePusey1996}
P.~N. Segr\`{e} and P.~N. Pusey.
\newblock {\em Phys. Rev. Lett.}, 77:771, 1996.

\bibitem{Loewen1993}
H.~L\"{o}wen, T.~Palberg, and R.~Simon.
\newblock {\em Phys. Rev. Lett.}, 70:1557, 1993.

\bibitem{Holmqvist2010}
P.~Holmqvist and G.~N\"{a}gele.
\newblock {\em Phys. Rev. Lett.}, 104:058301, 2010.

\bibitem{Lurio2000}
L.~B. Lurio, D.~Lumma, A.~R. Sandy, M.~A. Borthwick, P.~Falus, S.~G.~J.
  Mochrie, J.~F. Pelletier, M.~Sutton, Lynne Regan, A.~Malik, and G.~B.
  Stephenson.
\newblock {\em Phys. Rev. Lett.}, 84:785, 2000.

\bibitem{Westermeier2012}
F.~Westermeier, B.~Fischer, W.~Roseker, G.~Gr\"{u}bel, G.~N\"{a}gele, and
  M.~Heinen.
\newblock {\em J. Chem. Phys.}, 137:114504, 2012.

\bibitem{Martinez2011}
V.~A. Martinez, J.~H.~J. Thijssen, F.~Zontone, W.~van Megen, and G.~Bryant.
\newblock {\em J. Chem. Phys.}, 134:054505, 2011.

\bibitem{NaegeleVarena2013}
G.~N{\"a}gele.
\newblock Colloidal hydrodynamics.
\newblock In C.~Bechinger, F.~Sciortino, and P.~Ziherl, editors, {\em Physics
  of Complex Colloids - Proceedings of International School of Physics Enrico
  Fermi}, page 507. IOS Amsterdam; SIF Bologna, 2013.

\bibitem{Pusey1990}
P.~N. Pusey.
\newblock {\em in Liquids, Freezing and the Glass Transition, Les Houches
  Session LI}.
\newblock Elsevier, Amsterdam, 1990.

\bibitem{LaddWeitz:1995}
J.X.~Zhu A.J.C.~Ladd, H.~Gang and D.A. Weitz.
\newblock {\em Phys. Rev. E}, 52:6550, 1995.

\bibitem{Abade2010}
G.~C. Abade, B.~Cichocki, M.~L. Ekiel-Je\.{z}ewska, G.~N\"{a}gele, and
  E.~Wajnryb.
\newblock {\em J. Chem. Phys.}, 132:014503, 2010.

\bibitem{vanMegenSnook1988}
W.~van Megen and I.~Snook.
\newblock {\em J. Chem. Phys.}, 88:1185, 1988.

\bibitem{NaegeleMolphys:2002}
G.~N\"{a}gele, M.~Kollmann, R.~Pesche, and A.J. Banchio.
\newblock {\em Molec. Phys.}, 100:2921, 2002.

\bibitem{Haertl1992}
W.~H\"{a}rtl, H.~Versmold, W.~Wittig, and P.~Linse.
\newblock {\em J. Chem. Phys.}, 97:7797, 1992.

\bibitem{WagnerHaertl:2001}
J.~Wagner, W.~H\"{a}rtl, and H.~Walderhaug.
\newblock {\em J. Chem. Phys.}, 114:975, 2001.

\bibitem{Banchio2008}
A.~J. Banchio and G.~N\"{a}gele.
\newblock {\em J. Chem. Phys.}, 128:104903, 2008.

\bibitem{Heinen2010}
M.~Heinen, P.~Holmqvist, A.~J. Banchio, and G.~N\"{a}gele.
\newblock {\em J. Appl. Crystallogr.}, 43:970, 2010.

\bibitem{Heinen2011_dyn}
M.~Heinen, A.~J. Banchio, and G.~N\"{a}gele.
\newblock {\em J. Chem. Phys.}, 135:154504, 2011.

\bibitem{Heinen2011}
M.~Heinen, P.~Holmqvist, A.~J. Banchio, and G.~N\"{a}gele.
\newblock {\em J. Chem. Phys.}, 134:044532 and 129901, 2011.

\bibitem{SegrePusey1997}
P.N. Segr\`{e} and P.N. Pusey.
\newblock {\em Physica A}, 235:9, 1997.

\bibitem{PuseyPoon1997}
P.N Pusey, P.N. Segr\`{e}, O.P. Behrend, S.P. Meeker, and W.C.K. Poon.
\newblock {\em Progr. Colloid Polym Sci.}, 104:8, 1997.

\bibitem{vanMegenUnderwood:1989}
W.~van Megen and S.M. Underwood.
\newblock {\em J. Chem. Phys.}, 91:552, 1989.

\bibitem{vanMegenSchoepe:2017}
W~van Megen and H.J. Sch\"{o}pe.
\newblock {\em J. Chem. Phys.}, 146:104503, 2017.

\bibitem{Brands:1999}
A.~Brands, H.~Versmold, and W.~van Megen.
\newblock {\em J. Chem. Phys.}, 110:1283, 1999.

\bibitem{Lumma2000}
D.~Lumma.
\newblock {\em Phys. Rev. E}, 62:8258, 2000.

\bibitem{Joshi:2013}
R.G. Joshi, B.V.R. Tata, and J.~Brijitta.
\newblock {\em J. Chem. Phys.}, 139:124901, 2013.

\bibitem{Sigel:1999}
R.~Sigel, S.~Pispas, D.~Vlassopolous, N.~Hadjichristidis, and G.~Fytas.
\newblock {\em Phys. Rev. Lett.}, 83:4666, 1999.

\bibitem{CichockiFelderhof:1993}
B.~Cichocki and B.U. F.
\newblock {\em J. Chem. Phys.}, 98:8186, 1993.

\bibitem{CichockiFelderhof:1994}
B.~Cichocki and B.U. Felderhof.
\newblock {\em Physica A}, 204:152, 1994.

\bibitem{FuchsMayr:1999}
M.~Fuchs and M.R. Mayr.
\newblock {\em Phys. Rev. E}, 60:5742, 1999.

\bibitem{Banchio:2000}
A.~J. Banchio, G.~N\"{a}gele, and J.~Bergenholtz.
\newblock {\em J. Chem. Phys.}, 113:3381, 2000.

\bibitem{Pusey1978}
P.~N. Pusey.
\newblock {\em J. Phys. A}, 11:119, 1978.

\bibitem{Beenakker1983}
C.~W.~J. Beenakker and P.~Mazur.
\newblock {\em Physica A}, 120:388, 1983.

\bibitem{Mazur1984}
C.~W.~J. Beenakker and P.~Mazur.
\newblock {\em Physica A}, 126:349, 1984.

\bibitem{Verwey_Overbeek1948}
E.~J.~W. Verwey and J.~T.~G. Overbeek.
\newblock {\em Theory of the Stability of Lyophobic Colloids}.
\newblock Elsevier, New York, 1948.

\bibitem{Russel1981}
W.~B. Russel and D.~W. Benzing.
\newblock {\em J. Colloid Interface Sci.}, 83:163, 1981.

\bibitem{Denton2000}
A.~R. Denton.
\newblock {\em Phys. Rev. E}, 62:3855, 2000.

\bibitem{Alexander1984}
S.~Alexander, P.~M. Chaikin, P.~Grant, G.~J. Morales, P.~Pincus, and D.~Hone.
\newblock {\em J. Chem. Phys.}, 80:5776, 1984.

\bibitem{HansenLoewen2000}
J.-P. Hansen and H.~L\"{o}wen.
\newblock {\em Annu. Rev. Phys. Chem.}, 51:209, 2000.

\bibitem{Trizac2003}
E.~Trizac, L.~Bocquet, M.~Aubouy, and H.~H. von Gr\"{u}nberg.
\newblock {\em Langmuir}, 19:4027, 2003.

\bibitem{Pianegonda2007}
S.~Pianegonda, E.~Trizac, and Y.~Levin.
\newblock {\em J. Chem. Phys.}, 126:014702, 2007.

\bibitem{Torres2008}
A.~Torres, G.~Tellez, and R.~van Roij.
\newblock {\em J. Chem. Phys.}, 128:154906, 2008.

\bibitem{Colla2009}
T.~E. Colla, Y.~Levin, and E.~Trizac.
\newblock {\em J. Chem. Phys.}, 131:074115, 2009.

\bibitem{dosSantos2010}
A.~P. dos Santos, A.~Diehl, and Y.~Levin.
\newblock {\em J. Chem. Phys.}, 132:104105, 2010.

\bibitem{Denton2010}
A.~R. Denton.
\newblock {\em J. Phys.-Condes. Matter}, 22:364108, 2010.

\bibitem{Heinen2014}
M.~Heinen, T.~Palberg, and H.~L\"owen.
\newblock {\em J. Chem. Phys.}, 140:124904, 2014.

\bibitem{Boon2015}
N.~Boon, I.~G. Guerrero-Garcia, R.~van Roij, and M.~O. de~la Cruz.
\newblock {\em PNAS}, 112:9242, 2015.

\bibitem{Gruijthuisen:2013}
K.~van Gruijthuisen, M.~Obiols-Rabasa, M.~Heinen, G.~N{\"a}gele, and
  A.~Stradner.
\newblock {\em Langmuir}, 29:11199, 2013.

\bibitem{Philipse1989}
A.~P. Philipse and A.~Vrij.
\newblock {\em J. Colloid Interface Sci.}, 128:121, 1989.

\bibitem{Philipse1988}
A.~P. Philipse and A.~Vrij.
\newblock {\em J. Chem. Phys.}, 88:6459, 1988.

\bibitem{Thurn-Albrecht2003}
T.~Thurn-Albrecht, F.~Zontone, G.~Gr\"{u}bel, W.~Steffen,
  P.~M\"{u}ller-Buschbaum, and A.~Patkowski.
\newblock {\em Phys. Rev. E}, 68:031407, 2003.

\bibitem{Gapinski2009}
J.~Gapinski, A.~Patkowski, A.~J. Banchio, J.~Buitenhuis, P.~Holmqvist, M.~P.
  Lettinga, G.~Meier, and G.~N\"{a}gele.
\newblock {\em J. Chem. Phys.}, 130:084503, 2009.

\bibitem{ESRFwebpage}
\mbox{European Synchrotron Radiation Facility Beamline ID10}.
\newblock
  \newline\url{http://www.esrf.eu/UsersAndScience/Experiments/CBS/ID10}.

\bibitem{Schulz1939}
G.~V. Schulz.
\newblock {\em Z. Phys. Chem. B}, 43:25, 1939.

\bibitem{Zimm1948}
B.~H. Zimm.
\newblock {\em J. Chem. Phys.}, 16:1099, 1948.

\bibitem{Pusey1982}
P.~N. Pusey, M.~Fijnaut, H, and A.~Vrij.
\newblock {\em J. Chem. Phys.}, 77:4270, 1982.

\bibitem{Egelhaaf2004}
S.~U. Egelhaaf, V.~Lobaskin, H.~H. Bauer, H.~P. Merkle, and P.~Schurtenberger.
\newblock {\em Eur. Phys. J. E}, 13:153, 2004.

\bibitem{Heinen2012}
M.~Heinen, F.~Zanini, F.~Roosen-Runge, D.~Fedunov\'{a}, F.~Zhang, M.~Hennig,
  T.~Seydel, R.~Schweins, M.~Sztucki, M.~Antal\'{\i}k, F.~Schreiber, and
  G.~N\"{a}gele.
\newblock {\em Soft Matter}, 8:1404, 2012.

\bibitem{BernePecora1976}
B.~J. Berne and R.~Pecora.
\newblock {\em Dynamic Light Scattering: With Applications to Chemistry,
  Biology, and Physics}.
\newblock Wiley, New York, 1 edition, 1976.

\bibitem{Pedersen1997}
J.~S. Pedersen.
\newblock {\em Adv. Colloid Interface Sci.}, 70:171, 1997.

\bibitem{Akerlof1932}
G.~\AA{}kerl\"of.
\newblock {\em J. Am. Chem. Soc.}, 54:4125, 1932.

\bibitem{CRChandbook}
R.C. Weast and D.~R. Lide.
\newblock {\em CRC Handbook of Chemistry and Physics}.
\newblock CRC Press, Boca Raton, FL, USA, 56th edition, 1975.

\bibitem{Li1994}
W.~B. Li, P.~N. Segr\`{e}, R.~W. Gammon, J.~V. Sengers, and M.~Lamvik.
\newblock {\em J. Chem. Phys.}, 101:5058, 1994.

\bibitem{Wang2012}
R.~Wang, J.-q. Yao, Y.~Lu, Y.-p. Miao, X.-l. Zhao, and R.-q. Wang.
\newblock {\em Optoelectronics Letters}, 8:430, 2012.

\bibitem{Dhont1996}
J.~K.~G. Dhont.
\newblock {\em An Introduction to Dynamics of Colloids}.
\newblock Elsevier, Amsterdam, 1996.

\bibitem{Rogers1984}
F.~J. Rogers and D.~A. Young.
\newblock {\em Phys. Rev. A}, 30:999, 1984.

\bibitem{KimKarilla}
S.~Kim and S.~J. Karrila.
\newblock {\em Microhydrodynamics: Principles and Selected Applications}.
\newblock Butterworth-Heinemann, Boston, MA, 1991.

\bibitem{Makuch2012}
K.~Makuch and B.~Cichocki.
\newblock {\em J. Chem. Phys.}, 137:184902, 2012.

\bibitem{Makuch2015}
K.~Makuch.
\newblock {\em Phys. Rev. E}, 92:042317, 2015.

\bibitem{Jeffrey1984}
D.~J. Jeffrey and Y.~Onishi.
\newblock {\em J. Fluid Mech.}, 139:261--290, 1984.

\bibitem{NageleVarenna2013}
G.~N\"{a}gele.
\newblock Colloidal hydrodynamics.
\newblock In {\em Physics of Complex Colloids – Proccedings of the
  International School of Physics Enrico Fermi, edited by C. Bechinger, F.
  Sciortino and P. Ziherl. (2013)}.

\bibitem{Naegele1997}
G.~N\"{a}gele and P.~Baur.
\newblock {\em Physica A}, 245:297, 1997.

\bibitem{Pottier:2011}
N~. Pottier.
\newblock {\em Physica A}, 390:2863, 2011.

\bibitem{NaegeleDhont:1998}
G.~N\"{a}gele and J.K.G. Dhont.
\newblock {\em J. Chem. Phys.}, 108:9566, 1998.

\bibitem{SierouBrady2001}
A.~Sierou and J.F. Brady.
\newblock {\em J. Fluid Mech.}, 448:115, 2001.

\bibitem{BanchioBrady2003}
A.~J. Banchio and J.~F. Brady.
\newblock {\em J. Chem. Phys.}, 118:10323, 2003.

\bibitem{Ladd:1990}
A.J.C. Ladd.
\newblock {\em J. Chem. Phys.}, 93:3484, 1990.

\bibitem{Abade2011}
G.~C. Abade, B.~Cichocki, M.~L. Ekiel-Je\.{z}ewska, G.~N\"{a}gele, and
  E.~Wajnryb.
\newblock {\em J. Chem. Phys.}, 134:244903, 2011.

\bibitem{Wang2015}
M~Wang, M.~Heinen, and J.~F. Brady.
\newblock {\em J. Chem. Phys.}, 142:064905, 2015.

\bibitem{CichockiHinsen:1992}
B.~Cichocki and K.~Hinsen.
\newblock {\em Physica A}, 187:133, 1992.

\bibitem{Gapinski2014}
J.~Gapinski, G.~N\"{a}gele, and A.~Patkowski.
\newblock {\em J. Chem. Phys.}, 141:124505, 2014.

\bibitem{Gapinski2010}
J.~Gapinski, A.~Patkowski, and G.~N\"{a}gele.
\newblock {\em J. Chem. Phys.}, 132:054510, 2010.

\bibitem{Banchio1999}
A.~J. Banchio, G.~N\"{a}gele, and J.~Bergenholtz.
\newblock {\em J. Chem. Phys.}, 111:8721, 1999.

\bibitem{Segre1995}
P.~N. Segr\`{e}, O.~P. Behrend, and P.~N. Pusey.
\newblock {\em Phys. Rev. E}, 52:5070, 1995.

\bibitem{Genz1991}
U.~Genz and R.~Klein.
\newblock {\em Physica A}, 171:26, 1991.

\bibitem{Riest:2015}
J.~Riest, T.~Eckert, W.~Richtering, and G.~N\"{a}gele.
\newblock {\em Soft Matter}, 11:2821, 2015.

\bibitem{vanMegenSnook1981}
W.~van Megen and I.~Snook.
\newblock {\em J. Chem. Phys.}, 75:1682, 1981.

\bibitem{Tough:1986}
H.N.W.~Lewkkerkerker R.J.A.~Tough, P.N.~Pusey and C.van den Broeck.
\newblock {\em Molec. Phys.}, 59:595, 1986.

\bibitem{FuchsGaussian:1998}
M.~Fuchs, W.~G{\"o}tze, and M.R. Mayr.
\newblock {\em Phys. Rev. E}, 58:3384, 1998.

\bibitem{FlennerSzamel:2005}
E.~Flenner and G.~Szamel.
\newblock {\em Phys. Rev. E}, 72:031508, 2005.

\bibitem{RiestSALR:2015}
J.~Riest and G.~N\"{a}gele.
\newblock {\em Soft Matter}, 11:9273, 2015.

\bibitem{Shiba:workinprogress}
S.~Das, J.~Riest, R.G. Winkler, G.~Gompper, J.K.G. Dhont, and G.~N\"{a}gele.
\newblock accepted (2017), doi:10.1039/c7sm02019h.
\newblock {\em Soft Matter}.

\end{thebibliography}
\end{document}